MDPI

Article

# Measuring Price Risk Aversion through Indirect Utility Functions: A Laboratory Experiment

Ali Zeytoon-Nejad

School of Business, Wake Forest University, Winston-Salem 27109, NC, USA; zeytoosa@wfu.edu

**Abstract:** The present paper introduces a theoretical framework through which the degree of risk aversion with respect to uncertain prices can be measured through the context of the indirect utility function (IUF) using a lab experiment. First, the paper introduces the main elements of the duality theory (DT) in economics. Next, it proposes the context of IUFs as a suitable framework for measuring price risk aversion through varying prices as opposed to varying payoffs, which has been common practice in the mainstream of experimental economics. Indeed, the DT in modern microeconomics indicates that the direct utility function (DUF) and the IUF are dual to each other, implicitly suggesting that the degree of risk aversion (or risk seeking) that a given rational subject exhibits in the context of the DUF must be equivalent to the degree of risk aversion (or risk seeking) elicited through the context of the IUF. This paper tests the accuracy of this theoretical prediction through a lab experiment using a series of relevant statistical tests. This study uses the multiple price list (MPL) method, which has been one of the most popular sets of elicitation procedures in experimental economics to study risk preferences in the experimental laboratory using non-interactive settings. The key findings of this study indicate that price risk aversion (PrRA) is statistically significantly greater than payoff risk aversion (PaRA). Additionally, it is shown that the risk preferences elicited under the expected utility theory (EUT) are somewhat subject to context. Other findings imply that the risk premium (RP), as a measure of willingness to pay for insuring an uncertain situation, is statistically significantly greater for stochastic prices compared to that for stochastic payoffs. These results are robust across different MPL designs and various statistical tests that are utilized.

## 1. Introduction

Eliciting the degree of risk aversion is of crucial importance in psychology, finance, as well as economics. In particular, it is essential to the psychophysics of chance, financial decision making, and economic modeling. A highly prevalent approach to eliciting risk preferences in experimental economics is to use the *direct* utility function (DUF) for this purpose. However, the present paper uses an experimental design that employs the *indirect* utility function (IUF) to elicit risk preferences. Thereby, the degrees of PaRA and PrRA can be reasonably compared.

The duality theory (DT) in modern microeconomics suggests that the DUF and the IUF are dual of each other, meaning that, when one is known, it can be used to theoretically derive the other. For instance, if an IUF is known, then one can simply use Roy's identity to derive a system of Marshallian demand functions, and then substitute the Marshallian demand functions derived into the IUF to find the corresponding DUF. As a result, the DT implicitly suggests that the degree of risk aversion (or risk seeking) that a given (rational) subject exhibits in the context of the DUF must be equivalent to the degree of risk aversion (or risk seeking) elicited through the context of the IUF. However, the accuracy of this theoretical prediction remains an empirical question, which can be tested in a lab experiment. In light of this, one of the objectives of the present paper is to investigate whether this theoretical prediction (which is mathematically appealing) is confirmed by experimental evidence or not. This paper also examines whether there are potential behavioral interpretations for any gap that may be observed between the two approaches. The general approach of this study is to rely on elicitations that use payoff-based lottery choices (which are based on the DUF and uncertainty about payoffs) versus their equivalent price-based lottery choices (which are based on the IUF and uncertainty about prices).

This paper expands upon Zeytoon-Nejad et al. (2020) [1]'s short letter/note (which provided some preliminary results very briefly using other statistical analyses—i.e., random-effects ordered probit regression models) and elaborates on the experimental design, theoretical framework, and methodology of the related experiment, and provides additional complementary analyses on their laboratory experiment to address additional aspects of the experiment and introduce the additional findings and results of that experiment. In this paper, emphasis is placed on theoretical framework, experimental design, methodology, as well as checking the robustness and reliability of the results across all the methods used. In another paper (Zeytoon-Nejad, 2022) [2], a gender comparison of the findings is provided, explaining different risk attitudes and distinct behaviors under uncertainty across gender. In this study, a lab experiment was designed that enables researchers to elicit risk attitudes, measure the degrees of risk aversion, and estimate risk premiums through IUFs. To do so, the frameworks of three popular multiple price list (MPL) designs in the area of experimental economics were adopted, including Holt and Laury (2002) [3], Binswanger (1980) [4], and the certainty vs. uncertainty design, which has been applied in different forms by a number of scholars in the field (henceforth, the H&L, Bins., and CvU designs, respectively). These designs have been the subject of many experimental and empirical studies, but all of them have used these designs in the context of the DUF. However, in this experiment, all **three MPL designs** (i.e., three contexts) were used as elicitation procedures, each of which has **two versions** (i.e., two approaches)—a version with a DUF approach and another with an IUF approach. Accordingly, this experimental study has a 3 × 2 design. In short, the three MPL designs are as follows:

- **Holt and Laury** design (**H&L**)—has the advantage of "varying probabilities" (or "probability weighting"), which is an important feature of the expected utility theory (henceforth, the EUT).
- **Binswanger** design (**Bins.**)— has the advantage of "varying payoffs" (i.e., weighting payoffs), which is another important feature of the EUT.
- **Certainty vs. uncertainty** (**CvU**) design—has the advantage of investigating decision making under both "certainty vs. uncertainty", which is another important aspect of the EUT.

In this experiment, these six risk elicitation procedures were deliberately designed and calibrated such that, given the DT and the EUT, each should elicit the same degree of risk aversion exhibited by a given rational individual. In practice, each laboratory subject was exposed to both DUF and IUF frameworks, and within either of these two frameworks, he or she was asked to carry out the three tasks associated with the three MPL designs mentioned above. Therefore, it could be found out whether, and if so, to what extent, the elicited results differ between the two frameworks (DUF vs. IUF), as well as from one MPL design to another (H&L vs. Bins. vs. CvU). Furthermore, attempts were made to identify and explain the systematic differences among the degrees of risk aversion elicited from each approach and design.

The primary research questions of this experiment can be listed as follows: (1) How can one employ the IUF framework in order to directly measure the degree of price risk aversion that experimental subjects exhibit in the lab? (2) What type of risk attitudes do the subjects exhibit under uncertainty regarding the prices of goods they are to buy? (3) If risk aversion is the dominant risk attitude (which is a typical finding from lottery choice experiments of this sort conducted in the context of the DUF), to what extent are the subjects risk-averse? (4) How much are the subjects' risk premiums over the lotteries defined with stochastic prices? In other words, how large of a premium are the subjects willing to pay to set prices fixed ex ante (e.g., setting the price on a contract today and fixing it now, as opposed to leaving it to be determined tomorrow by the respective market forces in a stochastic way and thereby carrying out transactions under uncertain future market prices)? (5) How different are these results from those obtained from the DUF (with stochastic payoffs) for the same individual?

To find the answers to the above-mentioned questions, a 3 × 2 experimental design was employed. That is, each of the three MPL designs (H&L, Bins., and CvU) was examined in the context of the DUF as well as that of the IUF (the *direct* approach as well as the *indirect* approach). In fact, risk attitudes were evaluated by asking subjects to make a series of choices over lotteries that involved some degree of uncertainties, either with payoff odds or with price odds. For each of the six treatments, four independent sessions were carried out. The order of tasks in each session was randomly assigned to account for any potential order effects and learning effects. Numerous socio-demographic variables were also controlled for. The gender differences and socio-demographic characteristics of PaRA and PrRA are to be addressed and discussed in a separate paper, which is still under preparation. The subjects were students studying at North Carolina State University. Altogether, 88 students from a range of disciplines participated in the experiments, and the average payoff was USD 16.76 (including a USD 5 participation payment). All the subjects participating in the experiment conducted the tasks using the computers in the experimental economics laboratory of the Department of Economics at North Carolina State University. The popular experimental economics software zTree was employed for the purpose of this lab experiment.

The main findings of the study show that the vast majority of subjects are risk-averse, regardless of whether the elicitation approach is direct (through the DUF) or indirect (through the IUF). In fact, only few (less than 5%) of them exhibit risk-loving attitudes, and the rest are either risk-neutral (about 12%) or risk-averse (about 83%), averaged across the tasks. Although some economists argue that decision makers should be approximately risk-neutral for the low-payoff decisions (involving several dollars) that are typically encountered in the laboratory, the results of this study strongly conflict this view. In addition, surprisingly, the subjects exhibit statistically significantly greater degrees of risk aversion when faced with **random prices** (PrRA) compared to when faced with equivalent **random payoffs** (PaRA). This is a remarkable and thought-provoking result. More specifically, the findings indicate that the average of the estimated midpoint CRRAs is equal to 0.597 for PaRA (which implies a '**risk-averse**' attitude), while it is equal to 0.708 for PrRA (which implies a '**very risk-averse**' attitude). More interestingly, this result (i.e., **PaRA < PrRA**) is robust across all the MPL designs that were used, which indicates that the observed anomalies in the degrees of risk aversion exhibited by the subjects are quite systematic (consistent across designs and subjects), and, as such, can reasonably and convincingly be attributed to the nature of each approach (i.e., the inherently different risk preferences that subjects exhibit with respect to random prices and random payoffs).

Additionally, it is shown that risk elicitation results (risk attitudes and the degrees of risk aversion) elicited under the EUT are somewhat subject to context (i.e., the three MPL designs), and this result is consistent with the mainstream experimental literature that has revealed that context matters when results are generated under the EUT. For a good discussion of this topic, as an example, you can see Zhou and Hey (2017) [5]. However, our findings imply that the broadly defined "risk attitudes" (i.e., being risk-loving, risk-neutral, and risk-averse) elicited under the EUT are not subject to context to the same extent. Thus, a conclusion that can be drawn from our results is that the MPL elicitation method (which we call the context of elicitation) matters to the estimated "degree of risk aversion", but not much so to the broadly categorized "risk attitudes" elicited.

For the purpose of statistical hypothesis testing, a wide range of pertinent statistical tests are used, including the Wilcoxon signed-rank test, the Arbuthnott–Snedecor–Cochran sign test, and the two-sample T test for paired data, and the great majority of the mentioned statistical tests confirm that **PrRA is statistically significantly greater than PaRA**. This implicitly suggests that individuals, in general, have higher willingness to pay (WTP) for price-guaranteeing insurance premiums than those guaranteeing payoff quantities. It also indicates that risk-preference-related implications of the duality theory are rejected from a behavioral point of view, since experimental evidence shows that there is a systematic distance from rationality when subjects are exposed to random payoffs versus random prices.

The remainder of this paper will proceed as follows. Section 2 is devoted to explaining the methodology and theoretical considerations related to the subject matter of this study, in which the way of measuring risk aversion through the three popular MPL designs are explained, and afterwards, the notion of risk aversion in the context of the DUF is theoretically compared with that in the context of the IUF. In Section 3, our experimental design and procedures are introduced. Next, Section 4 describes the data and variables used in the analysis. Section 5 outlines the estimation strategy and procedures applied to elicit risk attitudes and measure the degree of risk aversion. After that, the results of the study and estimations are reported. In Section 6, the results are summarized, organized, and discussed. In Section 7, naturally, a conclusion will follow bringing the main points and major findings together and discussing plans for future research. Lastly, the paper will end with appendices to explain the data, designs, procedures, methods, and tests in greater detail (the dataset, appendices, and experiment instructions of this study are available at: https://zeytoonnejad.wordpress.ncsu.edu/my-research-2/, accessed on 16 June 2022).

## 2. Methodology and Theoretical Considerations

This section includes two sub-sections. Section 2.1 explains how risk aversion is measured through the three MPL risk elicitation designs discussed above, and also illustrates ways that the menu of lottery choices can be deliberately calibrated to be equivalent to each other for a given rational individual. Section 2.2 attends to the duality theory and its implications for risk attitudes and provides theoretical considerations for risk aversion in the DUF versus risk aversion in the IUF.

### 2.1. Measuring Risk Aversion through Three Popular MPL Risk Elicitation Designs

Following the same practice by Holt and Laury (2002) [3], risk attitudes were classified under the categories presented in Table 1. In fact, the risk aversion categories reported in this table were used to design the menu of lottery choices (i.e., the tasks). Then, the menus of lottery choices of the other two MPL designs (i.e., the CvU and Bins. designs) are calibrated such that the risk aversion intervals and the number of safe choices (for the CvU design) and the selected decision numbers (for the Bins. design) remain equivalent and correspond across the three MPL designs, given the EUT.

Table 1. Risk aversion classifications based on options chosen.

| Number of Safe Choices (For the HL and CvU Designs) | Selected Decision Number (For the Bins. Design) | Range of the Implied Coefficients of RRA for the CRRA Utility Function | Risk Attitude Classifications |
|---|---|---|---|
| 0–1 | 1 | $r < -0.95$ | Highly risk-loving |
| 2 | 2 | $-0.95 < r < -0.49$ | Very risk-loving |
| 3 | 3 | $-0.49 < r < -0.15$ | Risk-loving |
| 4 | 4 | $-0.15 < r < 0.15$ | Risk-neutral |
| 5 | 5 | $0.15 < r < 0.41$ | Slightly risk-averse |
| 6 | 6 | $0.41 < r < 0.68$ | Risk-averse |
| 7 | 7 | $0.68 < r < 0.97$ | Very risk-averse |
| 8 | 8 | $0.97 < r < 1.37$ | Highly risk-averse |
| 9–10 | 9 or 10 | $r > 1.37$ | Stay in bed (extremely risk-averse) |

**Note:** The implied CRRA coefficients apply to all the three MPL designs in the way outlined in the table. In fact, the task designs are deliberately calibrated and arranged such that these classifications hold true for all the three MPL designs.

**Note:** One may argue that the final results of this study could be sensitive to utility functional forms. However, in the literature, it has already been shown that the elicited degree of risk aversion is not very sensitive to utility functional form. Rather, it is more sensitive to the elicitation method (i.e., context). For instance, Zhou and Hey (2017) [5] used two expected utility and rank-dependent expected utility functionals, each of which combined with either a CRRA or a CARA utility function (i.e., four functional forms in total). Their findings indicate that the inferred level of risk aversion is more sensitive to the elicitation method rather than to the assumed-true preference functionals, and even less sensitive to the utility functional forms used. Additionally, Heinemann (2008) [6] showed that most subjects' behavior is consistent with CRRA. As a result, since the choice of the utility functional form is not of great concern for the research purposes, this paper adheres to the CRRA utility functional form, which has been the most commonly used and widely accepted functional form by scholars in the field.

**Note:** The expected payoffs and expected payoff differences for each of the corresponding price-based lotteries (which will be introduced and explained in Appendix C) remain the same as those of their corresponding payoff-based lotteries, which are reported above.

For the purpose of the utility functional form, the literature of experimental economics usually assumes constant relative risk aversion (CRRA). This functional form is assumed primarily for its computational convenience, theoretical support, robust predictions, and mathematical tractability. With CRRA for the monetary amount *x*, the utility function is defined as $u(x) = x^{1-r}/(1-r)$ for $x > 0$ and $r \neq 1$, and $u(x) = log(x)$ for $x > 0$ and $r = 1$. This utility function specification implies the risk-loving preference for $r < 0$, the risk-neutral preference for $r = 0$, and the risk-averse preference for $r > 0$ Following Arrow (1965) [7] and Pratt (1964) [8], the measure of risk aversion in this study is the Arrow–Pratt measure of Relative Risk Aversion (RRA), aka the Coefficient of Relative Risk Aversion (CRRA), which is defined as $R(x) = -x \frac{u''(x)}{u'(x)}$, where $u'(x)$ and $u''(x)$ denote the first and second derivatives of the utility function with respect to *x*, respectively. Assuming a CRRA utility function, it is possible to calculate an interval estimate of the CRRA (that is, *r*), as achieved by Holt and Laury (2002) [3]. For instance, the CRRA of a subject that picks '*n*' times option A before switching to option B must satisfy a set of two equations of the following form:

$$P_{A1}^{(n)} \cdot u(x_{A1}^{(n)}) + P_{A2}^{(n)} \cdot u(x_{A2}^{(n)}) \geq P_{B1}^{(n)} \cdot u(x_{B1}^{(n)}) + P_{B2}^{(n)} \cdot u(x_{B2}^{(n)}) \quad \text{for row } n \quad (1)$$

$$P_{A1}^{(n+1)} \cdot u(x_{A1}^{(n+1)}) + P_{A2}^{(n+1)} \cdot u(x_{A2}^{(n+1)}) \leq P_{B1}^{(n+1)} \cdot u(x_{B1}^{(n+1)}) + P_{B2}^{(n+1)} \cdot u(x_{B2}^{(n+1)}) \quad \text{for row } n+1 \quad (2)$$

By plugging Equation (1) into Equation (2) and solving for an *r* interval, one can obtain the subject's implied CRRA interval. Almost the same set of EUT equations must hold true for the other two MPL designs (the CvU and Bins. designs) with minor differences; in the CvU design, one side of each equation is a certain payoff, and in the Bins. design, each row makes an equation with its next row, as opposed to the H&L design in which each row has an equation of its own.

Appendix G presents the three MPL designs used in this study. As shown in the appendix, to compute the degree of risk aversion through the H&L design, subjects are asked 10 times to choose between Option A (a relatively less risky option) and Option B (a relatively more risky option). In the CvU design, subjects are asked 10 times to choose between Option A (a certain option) and Option B (an uncertain option). In these two designs, an extremely risk-averse subject will always prefer Option A over Option B, whereas an extremely risk-seeking subject will always prefer Option B over Option A. The row number at which a subject switches from Option A to Option B implies an interval estimate of the degree of risk aversion. This is because the row number of switching (i.e., the switching point) indicates the number of safe choices and thereby the degree of risk aversion. Therefore, the later a subject switches from Option A to Option B, the more risk-averse the subject will be. The Bins. design is somewhat different from the H&L and CvU designs in that the subject is supposed to choose only one time and makes only one choice out of ten options. In other words, the subject makes only one decision, in which they face ten options of uncertain payoffs (i.e., ten lotteries). As the subject moves down the menu, the magnitudes of the two possible payoffs of the choice listed in each row become closer to one another (i.e., they become less risky). In all the three designs, subjects' payoffs will depend on their choices and the payoff odds of the lotteries. For more information on the details and subtleties of each design, please see Appendix G, which provides the full instructions of the experiment.

*2.2. Risk Aversion in Direct Utility Function vs. Risk Aversion in Indirect Utility Function*

In his remarkable book entitled "*Duality in Modern Economics*", Cornes (2008) [9] attests that "dual arguments have, in recent years, become standard tools for analysis of problems involving optimization by consumers and producers". Nonetheless, dual arguments have not been discussed adequately in the context of the elicitation of risk attitudes, nor have they been sufficiently studied through the use of experimental methods yet. Accordingly, the present paper attempts to fill the two aforementioned gaps in the respective literature.

Duality, as a mathematical concept, is a vastly extensive subject matter. Hence, introducing all of its technical aspects in detail goes beyond the scope of the present paper. Rather, this section of the paper aims to provide a brief discussion of the theory of duality from a microeconomic perspective, and focusses more on the application of this theory in consumer choice, decision making under uncertainty, and analyzing choices under conditions of risk. For a more extensive discussion of various aspects of the duality theory in relation to the consumer theory, please see study by Moosavian (2016) [10] and Moosavian el al. (2018) [11] which visually decode the wheel of duality in consumer theory. For an extensive discussion of different components of the duality theory in relation to the production theory, you can see Naumenko and Moosavian (2016) [12]. These two papers present simple, clear, and holistic explanations of the components of dual arguments in economics. Additionally, Cornes (2008) [9] and Chambers (1988) [13] provide more extensive and detailed discussions of the theory of duality in economics.

The Direct Utility Function (henceforth, DUF) expresses utility as a function of quantities of real goods and/or payoffs. That is, the arguments of DUF are all quantities. As such, it takes the functional form of $U(q)$, where $q = (q_1, q_2, \ldots q_n)$ can be a vector of quantities. By contrast, the Indirect Utility Function (henceforth, IUF) is defined as the maximum utility that can be attained given a monetary budget and goods prices.[1] Therefore, it takes the functional form of $V(P, M)$, where $P = (P_1, P_2, \ldots P_n)$ can be a vector of prices and $M$ is budget or endowment. Indeed, a consumer's IUF shows the consumer's maximal attainable utility when faced with a vector of prices of goods and the consumer's budget or endowment. Thus, IUF echoes both the consumer's preferences as well as market conditions.

As explained by Moosavian (2016) [10], according to the duality theory (DT) in modern economics, the DUF and the IUF are dual of each other. Simply put, when one function is the dual function of another, it practically means that one function can be derived from the other. This is only a simple working definition of a dual function. From a mathematical point of view, there are more subtle and technical aspects to the definition of a dual function. To gain more information on a formal mathematical definition of a dual function, you can see mathematical textbooks on the theory of duality and convex optimization. For example, see Boyd and Vandenberghe (2004) [14]. The following two optimization problems show why the DUF and the IUF are called duals of each other.

$$U(q) \equiv \min_{P}\{V(P, M) | P.q \geq M\}$$

$$V(P, M) \equiv \max_{q}\{U(q) | P.q \leq M\}$$

As Moosavian (2016) [10] elaborates, given an IUF, one can use Roy's identity to derive a system of Marshallian demand functions (henceforth, MDFs) and then substitute the MDFs back into the IUF to find the corresponding DUF. On the contrary, given a DUF, one can maximize the DUF subject to the budget constraint at hand and use Lagrangian to arrive at the system of the MDFs and then susbtitute this derived demand system into the DUF and simplify terms to obtain the corrsponding IUF of interest.

Another way to make this transition is to utilize the Hotelling–Wold identity. In practice, one can first use the Hotelling–Wold identity to obtain a system of Hotelling-style inverse demand functions (HIDFs) from the DUF. This will result in a system of equations which expresses normalized prices as functions of quantity bundles. Next, this system is inverted and its price normalization needs to be undone so that we can transition from HIDFs to MDFs, and then substitute them back into the DUF so as to end up with the IUF. Zeytoon (2016) [15] has visually demonstrated all of these transitions.

It is important to note that preferences are situated in the DUF and the IUF. Each of these functions is essentially a single function containing all preferences over the commodities of interest. They are in fact an abstract form of preferences. Figure 1 briefly and visually summarizes all the relationships introduced above. It also provides all the operations, equations, identities, and lemmas that enable us to make the aforementioned transitions.

---

[1]. IUF is called indirect because consumers usually think about their preferences in terms of what they consume rather than prices and income.

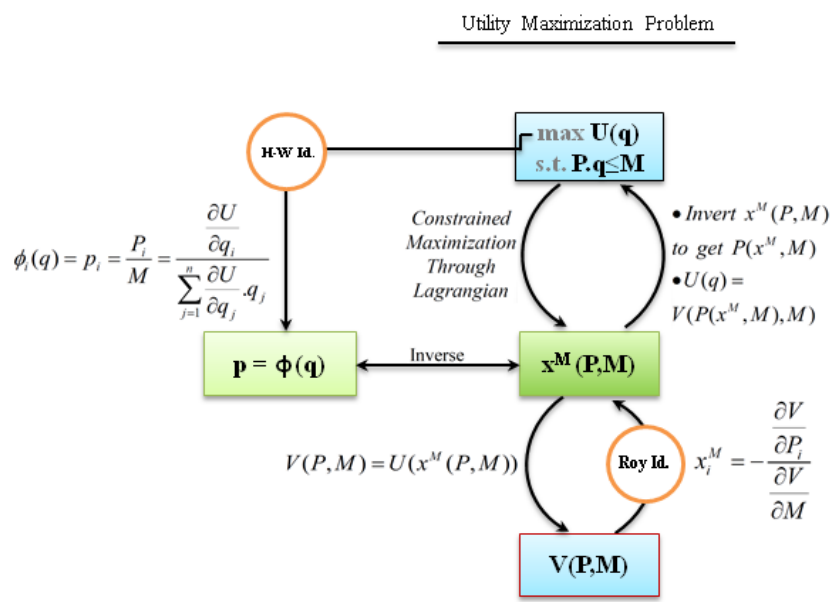

**Figure 1.** Different components of the Wheel of Duality in the utility maximization problem in the consumer theory. Adopted from Moosavian (2016) [10].

The operations, equations, identities, and lemmas needed to make the mentioned transitions are as follows: Lagrangian, mathematical substitution, mathematical inversion, price normalization (through dividing prices by income), Hotelling–Wold identity (labeled as H-W Id. in the visual), and Roy's identity (labeled as Roy's Id. in the visual). Appendix A provides the full version of the Wheel of Duality (WOD) in the consumer theory. For a full list of the symbols and notations employed in the visual WOD, please see Appendix B. A more detailed version of the lower-left portion of the above figure can be set up as follows.

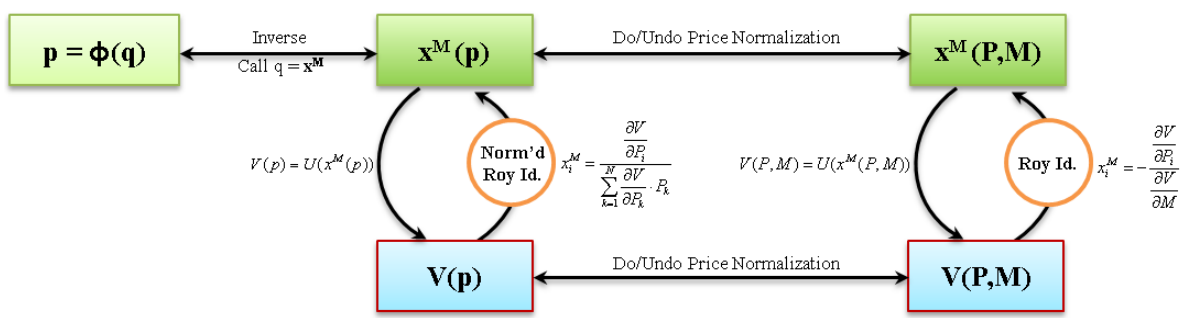

**Figure 2.** A more detailed version of the lower-left portion of the Wheel of Duality in the utility maximization problem. Adopted from Moosavian (2016) [10].

where $x^M(p)$ is in fact $x^M\left(\frac{P}{M}, \frac{M}{M}\right) = x^M(p, 1) = x^M(p)$ which is the MDF with normalized prices, and $V(p)$ is indeed $V\left(\frac{P}{M}, \frac{M}{M}\right) = V(p, 1) = V(p)$ which is the IUF with normalized prices. As shown in the above figure, the transition from $V(p)$ to $x^M(p)$ is made through the normalized version of the Roy's identity (labeled as Norm'd Roy Id. for short in the visual).

Now, consider the DUF and normalized IUF with only one argument in these functions, i.e., payoff for the former and widget price for the latter. A typical DUF is concave and takes the following form on the left, and a typical (normalized) IUF is convex and takes the following form on the right, as depicted in Figure 3:

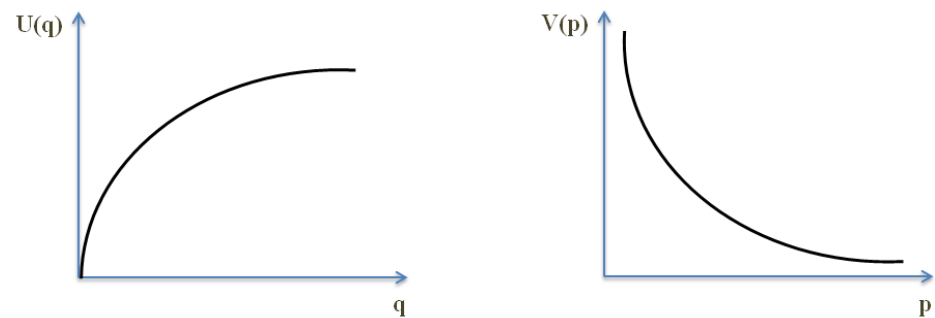

**Figure 3.** Direct utility function (DUF) versus indirect utility function (IUF). Source: Author's own drawing

The results of the duality theory which are often discussed for typical utility functions (which are defined over consumption bundles) in economics can also apply to Bernoulli-style utility functions (which represent preferences over sure monetary outcomes), and also apply to von-Neumann–Morgenstern (vNM) utility functions (which represent preferences over lotteries of monetary outcomes). In fact, the argument of the vNM-style utility function is the same argument as that of the Bernoulli-style utility function, and the primary difference between the two lies in the notion of probability weighting. Indeed, if one only considers degenerate lotteries, in which case the probabilities are either 0 or 1, then the vNM utility function and Bernoulli utility function coincide.

The indirect utility function, i.e., $V(P)$, takes the value of the maximum utility that can be attained by spending the budget $M$ on a good with the price of $P$. Accordingly, the DT implicitly suggests that the degree of risk aversion (or risk seeking) that a given rational subject exhibits in the context of the DUF must be equivalent to the degree of risk aversion (or risk seeking) elicited through the context of the IUF. However, the accuracy of this theoretical prediction remains an empirical question, which can be tested through a lab experiment. This paper is to investigate whether this theoretical prediction is verified by experimental evidence or not. Potential interpretable behavioral aspects to any gap that may be observed between the two approaches are examined. To this end, the methodology of this paper relies on elicitations that use payoff-based lottery choices (which is based on the DUF and uncertainty about payoffs) versus their equivalent price-based lottery choices (which is based on the IUF and uncertainty about prices).

The next section explains this approach in greater detail.

## 3. Experimental Design and Procedures

### 3.1. An MPL Design to Elicit Risk Attitudes in the Context of the IUF

In order to be able to compare the degree of the 'price' risk aversion (PrRA) with that of the 'payoff' risk aversion (PaRA) accurately and precisely, two inherently and theoretically equivalent menus of choice (or tasks) were needed to be considered for each MPL design: one with uncertainty about 'prices' and another with uncertainty about 'payoffs', so that we could interpret the potential differences in the elicited degrees of risk aversion as pure differential responses to uncertainty about 'prices' and uncertainty about 'payoffs'. To achieve this goal, this study took advantage of the duality theory in modern microeconomics and the sequence of the operations reported in Figures 1 and 2. Following the concepts and components of the duality theory associated with the DUF and the IUF, equivalent price-based versions of the three MPL designs of interest in this paper were designed, including the H&L, CvU, and Bins. designs. In simple words, the methodology of this research study relies on elicitations that use payoff-based lottery choices (based on the DUF and uncertainty about payoffs) versus their equivalent price-based lottery choices (based on the IUF and uncertainty about prices). By doing so, an experiment was designed, enabling us to elicit risk attitudes, measure the degrees of risk aversion, and estimate risk premiums by using the framework of the IUF.

As briefly discussed in the previous section, a consumer's IUF can be computed from his or her DUF by first computing the most preferred affordable bundle, represented by MDFs by solving the utility maximization problem, and next computing the utility that the consumer derives from that bundle. Accordingly, price-based MPL designs (i.e., tasks) were developed and calibrated by following the steps stated above, and by setting a budget constraint (with an endowment of USD 15), deriving a Marshallian demand, and finding and normalizing the IUF. Appendix G provides the instructions of the experiment and includes all the six choice menus associated with the mentioned six designs. Appendix C shows how the payoff-based lotteries are equivalent to price-based lotteries for the case of the H&L design. In this appendix, the first table corresponds to the framework of the DUF, the second table corresponds to the framework of the MDFs, and the third table corresponds to the framework of the IUF. The three tables presented in Appendix C illustrate the equivalence of the price-based lotteries and payoff-based lotteries for the H&L design. Similar tables can easily be derived and provided to illustrate the equivalence of the price-based lotteries and payoff-based lotteries for the other two MPL designs (i.e., the CvU and Bins. designs).

The next section explains the main approach and design in greater detail.

### 3.2. Experimental Design

As explained earlier, six equivalent risk elicitation designs are adopted and calibrated in such a way that, given the EUT and the DT, each should elicit the same degree of risk aversion exhibited by a given rational individual, although the designs differ in form, i.e., in terms of their approach (i.e., DUF vs. IUF) and their MPL designs (i.e., H&L, Bins., and CvU). In other words, these six elicitation procedures are theoretically equivalent, given the EUT and the DT. Hence, one can now investigate and test whether experimental evidence supports the equivalence of PaRA and PrRA, which is suggested by the DT. Moreover, one can examine and test whether experimental evidence supports the equivalence of the degrees of risk aversion elicited through each of the three MPL designs, as suggested by the EUT. Investigating these two matters can jointly enable us to attain the possibility of a robustness check across designs, which otherwise would have been impossible to attain. This way, the robustness of results can be checked with respect to a change in the contextual design (i.e., H&L, Bins., and CVU), as well as a change in approach (i.e., DUF vs. IUF). Accordingly, any significant differences across MPL designs can be interpreted as the fact that the EUT results are sensitive to context, and any significant differences across the approach can be interpreted as the fact that the suggestions made by the DT concerning the equivalency of risk attitudes elicited through either the DUF and the IUF do not hold true, in the sense that the degree of PrRA is not necessarily the same as the degree of PaRA.

In this lab experiment, subjects are presented with a set of lotteries. They play several games of chance, each involving a series of choices between two or more options. All of the games of chance involve odds, which work differently with different games. Understanding the odds is key to understanding the choices that the subjects should make.

A "task" is a table listing a set of lotteries. The subjects are asked to make their choices using the information provided in the table and report their decisions. The instructions for each particular task explain the related task in detail. In total, there are two types of tasks. In the first type of task (**Tasks 1–3**), the odds are called "**payoff odds**". In these tasks, the subjects choose between options shown to them on the computer screen. Payoff odds determine the payoff that the subjects will receive depending on their choices. This game of chance involves differing odds regarding whether the subjects receive a higher or lower "payoff".

For each of these three tasks, subjects are shown a set of lotteries on the computer screen. There are 10 rows for each task in total. Each row is a potential decision to be made.

Each lottery has its own possible payoffs. The rules and lotteries in each of these tasks are different. The instructions for each of the tasks are provided right before its related task. The most important thing to know for the subjects in these tasks is that they deal with "payoffs" in these tasks, i.e., "monetary payments" paid in cash as they leave the lab. They will choose between the options shown to them on the computer screen. Payoff odds determine the payoff they will receive depending on their choices. This game of chance involves differing odds regarding whether they will receive a higher or lower "payoff". As any one of the tasks could be the task that counts for the purpose of the payoff payment, they need to treat each task as if it could be the one that determines their payoff. What they earn will be their payoff.

In <u>the second type of task</u> (**Tasks 4–6**), the odds are called "**price odds**". In these tasks, the subjects buy and sell widgets. The price odds determine the price that the subjects will buy widgets at. This game of chance involves differing odds regarding whether the subjects will buy at a higher or lower "price".

Unlike Tasks 1–3, in Tasks 4, 5, and 6, subjects are faced with uncertain "prices". For each of these tasks, the computer gives the subjects a set of lotteries on a window. The windows show 10 rows (decisions). Each lottery has its own possible final payoffs, and the rules and lotteries in each of these tasks are different. The detailed instructions for each of the tasks are provided right before the related task. The most important thing to know for subjects in these tasks is the fact that they are dealing with uncertain "prices". In other words, in these three tasks, the odds are what we call "price odds". In these tasks, they will buy and sell widgets. In other words, they will buy widgets at some "uncertain buying prices" and will sell them to the experimenter at some "certain selling prices". Price odds determine the price that the subject will buy widgets at. This game of chance involves differing odds regarding whether the subject buys at a higher or lower "price". When subjects sell the widgets that they have already bought, they will receive the "monetary payments" that they will be paid in cash as they leave the lab. They choose between options shown to them on the computer screen. It is crucial for the subjects to know and remember that, in these three tasks, "buying prices" are uncertain. Price odds finally determine the payoff that subjects will receive, depending on their choices and chances. To determine their final payoffs, a six-sided die will eventually be rolled to select one of the six tasks for their payment at random.

Under this experimental design, several potential effects are addressed to the extent possible, including the "incentive effect", the "income effect", the "wealth effect", the "scale effect", the "endowment effect", the "learning effect", the "order effect", the "selection bias/effect", and the "fixed effects", so that any potential differences observed in the results of the six elicitation designs can purely and accurately be attributed to the phenomenon under study and can answer the research questions of interest in this study.

As for the incentive effect, the subjects are provided with inducement (i.e., real cash) to ensure that they are induced when making their decisions. Concerning the income effect, subjects are paid only once for one of the tasks chosen at random at the end of the experiment. Thus, there is no potential for any income effect. Additionally, this way of inducement and payment raises no concern over any potential accumulation of payoffs, thereby allowing us to hold wealth constant.

The experimental subjects are also asked about their own income and their family's income in a questionnaire, somehow accounting and controlling for these variables and their potential effects. According to Heinemann (2008) [6], "there is a long tradition in distinguishing two versions of expected utility theory: Expected Utility from Wealth (EUW) versus Expected Utility from Income (EUI)". As he puts it, "EUW assumes full integration of income from all sources in each decision and is basically another name for expected utility from consumption over time. EUI assumes that agents decide by evaluating the prospective gains and losses associated with the current decision, independent from initial wealth". Some believe that the degree of risk aversion is sensitive to assumptions about the wealth in the subjects' utility functions.

In order to avoid the payoff scale effect, all the menus of choice for the three MPL designs are developed such that the overall expected payoffs from the three payoff-based MPL designs are very close to each other. Needless to say, each price-based MPL design is exactly equivalent to its corresponding payoff-based MPL design in terms of its expected payoffs. Regarding the endowment effect, it should be noted that there is no possibility for the so-called endowment effect, since the subjects are explicitly told several times that they cannot take the endowment with them outside the lab, and that they have to buy the widgets using the endowment. In order to avoid any potential learning effect, subjects are not shown the outcome (payoff) of each task at the end of each task while they are conducting the tasks. In fact, they are shown the outcomes (payoffs) at the end of the experiment when they observe all the payoffs associated with each task, as well as the payoff selected at random for the purpose of payment.

In order to avoid any potential order effects (including both 'fatigue effect' and 'practice effect'), each subject sees the six tasks in a random order set by the computer. For example, one subject might be assigned to complete Task 6 first, and then Task 4 afterwards, and so on, while another subject might be assigned to complete Task 2 first, and then Task 1 afterwards, and so on, which are not necessarily the same as the order listed on the table of contents of the instructions. Everybody in the pool of subjects participated in all the six tasks, so there is no selection bias. Finally, we consider the fixed effects by making within-subjects comparisons, in which design the same group of subjects serves in all the six treatments. Therefore, there is no concern about subjects not being the same individuals.

In the next section, the experimental procedures are explained.

*3.3. The Experiment Procedures*

In this experiment, a 3 × 2 experimental design was used, in that each of the three MPL designs (H&L, CvU, and Bins.) was examined once in the context of the DUF and again in the context of the IUF. The experimental subjects were presented with a menu of choices that permits measurement of the degree of risk aversion, and enabled us to compare the behavior under uncertainty regarding payoffs and uncertainty about prices. Before conducting the actual experiment, the experiment was pre-tested and pilot-tested several times to ensure that all the aspects of the experiment were in order and the instructions were entirely understandable to typical subjects. Afterwards, for each of the six treatments, four independent sessions were carried out. The order of tasks for each subject in every session was randomly assigned to account for the potential order effect and the learning effect. Numerous demographic variables were also controlled for, such as age, gender, class status, college, major, financial independency, family income, etc. A full list of these variables is provided in Appendix E.

The subjects were students studying at North Carolina State University. Altogether, 88 students from a range of disciplines participated in the experiments, and the average payoff was USD 16.76 (including a USD 5 participation payment), with the lowest payoff being USD 5.60, and highest payoff being USD 28.08. Each session lasted approximately 75 min, with the first 15–20 min used for instructions. In order to make the two types of the tasks more understandable and salient to all the subjects, they were presented with two illustrative visuals describing the steps involved in the process of each type of task, along with the hard copies of the instructions for every subject, and pencils and calculators if requested. All the subjects participating in the experiment conducted the tasks using the computers in the experimental economics laboratory of the Department of Economics at North Carolina State University.

The popular experimental economics software zTree was employed for this lab experiment. In the zTree code, it was impossible for subjects to proceed with tasks if they made illogical and unreasonable answers (i.e., switching back and forth between Options A and B, which is not economically rational) by giving an error message which informed the subject why their decision was not rational. In total, only three subjects received this error message, and after understanding why their decisions were not rational, they reconsidered and corrected their answers.

After finishing the six treatments, the subjects answered the socio-demographic questionnaire. Finally, they were shown all the payoffs associated with each treatment using the computers, as well as the payoff selected at random for the purpose of payment at the end of the experiment. In the end, they were paid according to their randomly selected payoffs as they left the experimental laboratory.

The next section describes the data used in this study.

## 4. Data and Variables

This paper uses a dataset of 88 students studying at North Carolina State University (NCSU). The subjects were recruited randomly (conditional on being a student at NCSU) through an online recruitment system of the Experimental Economics Laboratory of NCSU. The subjects were between 19 and 28 years old in 2018. They were mostly (94%) from the US, and the others (6%) were from four other countries. Among the ones from the US, they were mostly (81%) from the state of North Carolina and the rest (19%) were from 12 other US states.

The main variables of interest in this study were the number of safe choices (for H&L and CvU) and the selected decision number (for Bins). In fact, given the EUT, these numerical values enable us to infer the degree of risk aversion for each subject with respect to the payoff (PaRA) as well as the price (PrRA). Other variables of interest include a set of socio-demographic variables, pointed out in the previous section, which can help identify some independent variables, thus explaining the degree of risk aversion.

Appendix F shows the full dataset of individual lottery choice decisions along with risk aversion classifications based on options chosen. This appendix also includes the questions of the socio-demographic questionnaire along with its numerical codes and the related demographic dataset.

In the next section, estimation results are provided.

## 5. Estimation and Results

### 5.1. Payoff Risk Aversion versus Price Risk Aversion

As reported in Table 1 and explained in the previous section, the main variables of interest in this study are the number of safe choices (in the H&L and CvU designs) and the selected decision number (in the Bins. design). These numbers are used to indicate the degree of risk aversion. In fact, these numerical values infer the degree of risk aversion for each subject with respect to payoff uncertainty (PaRA) in Tasks 1-3, as well as with respect to price uncertainty (PrRA) in Tasks 4–6. For a rational person, these choices must be equivalent in the way outlined in Table 1.

Figure 4 demonstrates the scatterplot matrix of the choices made by the subjects. One advantage of displaying choices in a scatterplot matrix of this form is that one can easily and instantly see the general pattern and distribution of risk attitudes by looking at the scatterplot matrix. As reported in Table 1, a risk-neutral subject can make four safe choices (in the H&L and CvU designs) or four pick decision numbers (in the Bins. design). Any selected number greater than four would imply a risk-averse attitude and any selected number smaller than four would imply a risk-loving attitude. Figure 4 displays the distributions of choices for each of the six treatments. As shown in this figure, the vast majority of subjects are risk-averse, regardless of whether the elicitation approach is direct (through the DUF) or indirect (through the IUF), as evidenced by the fact that most of the dots lie above (for vertical axes) and to the right of the number four (for horizontal axes) for all tasks. In fact, only few (less than 5%) of them exhibit risk-loving attitudes, and the rest are either risk-neutral (about 12%) or risk-averse (about 83%), averaged across the tasks. It is important to note that many of the dots, especially those representing choices and decisions greater than four, overlay and coincide with each other. As a consequence, the actual frequency of them is not observable on this scatterplot matrix. Therefore, one can further attend to these choices by plotting their histograms and kernel densities which will better represent the frequency of the choices made.

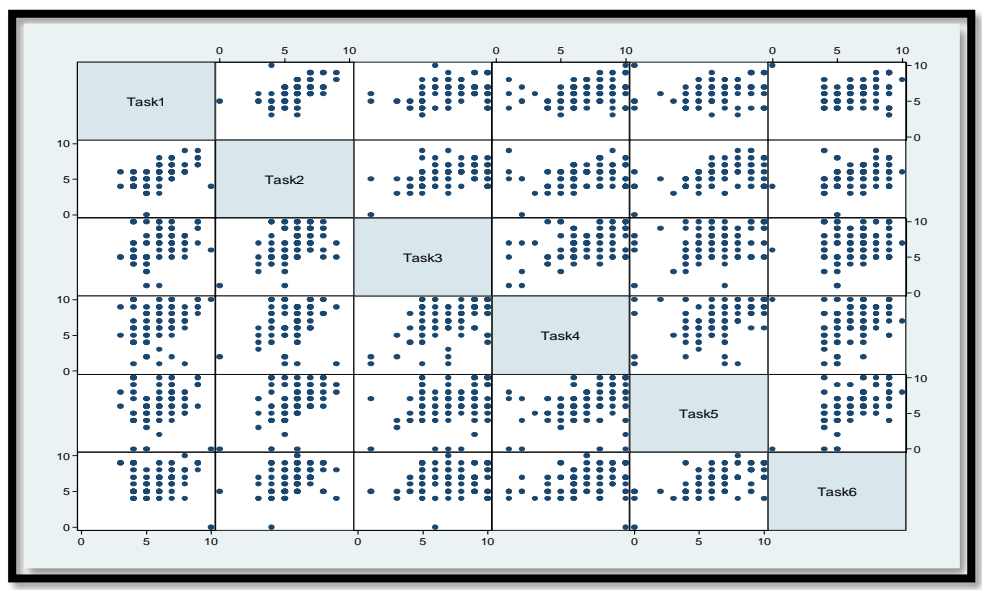

**Figure 4.** The scatter plot matrix of the choices made by the subjects.

Figure A2 in Appendix D depicts almost the same information as above but in terms of elicited midpoint CRRAs. In this case, the CRRA utility function specification implies the risk-loving attitude for CRRA < 0, the risk-neutral attitude for CRRA = 0, and the risk-averse attitude for CRRA > 0. As shown below, most of the subjects exhibit CRRA > 0.

An additional advantage of demonstrating choices in a scatterplot matrix form is that one can readily and visually compare the results from each payoff-based task (Tasks 1, 2, and 3) with the results from its corresponding price-based task (Tasks 6, 5, and 4, respectively). Moreover, one can visually notice differences in the results across the three MPL designs by comparing the PaRA results from Tasks 1, 2, and 3 with each other, and comparing the PrRA results from Tasks 4, 5, and 6 with each other. However, one disadvantage of exhibiting choices in a scatterplot form is that many of the dots may coincide with each other, hiding the actual frequency of the choices and decisions made by the subjects. To overcome this shortcoming, the respective histograms and kernel densities can be plotted, which will better represent the actual frequency of the choices and decisions made. Figure 5 exhibits the histograms and kernel densities of the switching points and decision numbers selected by the subjects, which represent the distributions of the degrees of risk aversion. In this figure, a diagram of each price-based task is placed under that of its corresponding payoff-based task for the sake of simplifying the comparison. In fact, the three diagrams on the top represent the frequencies of the degrees of PaRA, and the three diagrams on the bottom represent the frequencies of the degrees of PrRA. (The CRRA version of this figure is provided in Appendix D.)

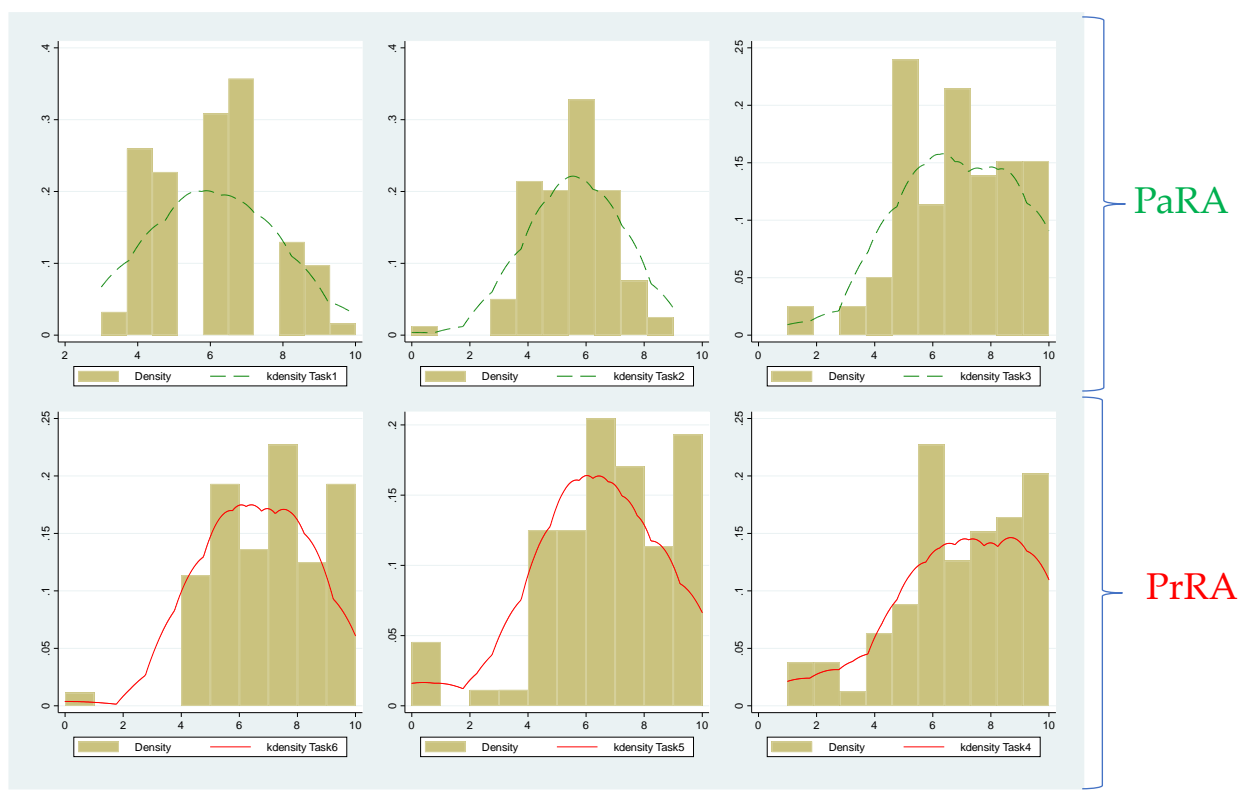

**Figure 5.** Histograms and kernel densities of the switching points and choice numbers selected by the subjects, which represent the distributions of the degrees of risk aversion.

In order to make the comparison of PrRA and PrRA even easier, Figure 6 provides a three-panel diagram, in which each pair of related kernel densities (i.e., Task 1 with 6, Task 2 with 5, and Task 3 with 4) is placed on the same diagram. Therefore, the relative frequencies of choices in each design can be easily compared with those of its corresponding design.

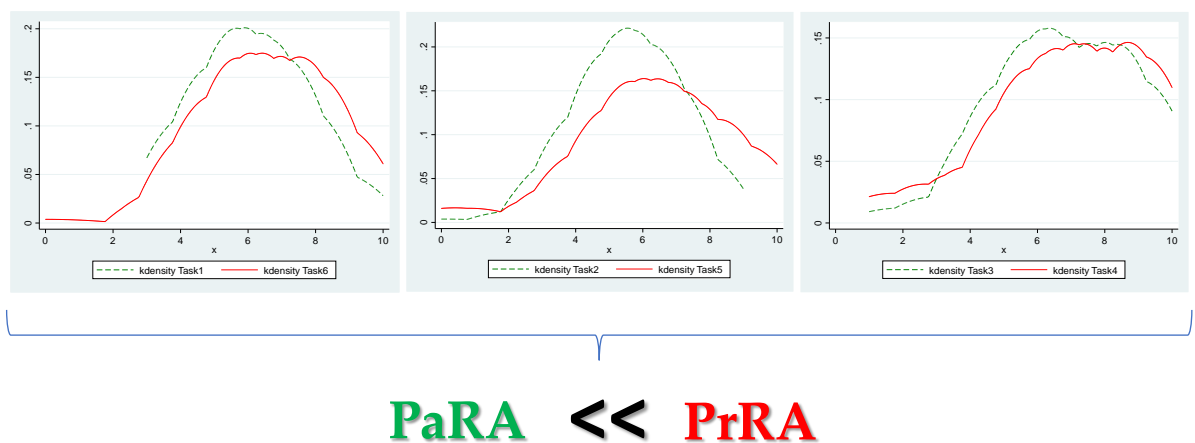

**Figure 6.** Kernel densities of the switching points and choice numbers selected by the subjects in corresponding designs. The dashes green lines represent the distributions of the degrees of PaRA and the solid red lines represent the degree of PrRA.

As demonstrated in Figures 5 and 6, the subjects exhibit considerably greater degrees of risk aversion when faced with **random prices** (PrRA) compared to when faced with **random payoffs** (PaRA). This observation can easily be made if one pays attention to the fact that in all the three pairs of designs, the red kernel density representing the degree of PrRA lies to the right of its green counterpart that represents the degree of PaRA. This is a remarkable and thought-provoking result. More specifically, the findings indicate that the average of the estimated midpoint CRRAs is equal to <u>0.597 for PaRA</u> (which implies the '**risk-averse**' attitude), while it is equal to <u>0.708 for PrRA</u> (which implies the '**very risk-averse**' attitude).

More interestingly, this result (i.e., **PaRA < PrRA**) is robust across all the MPL designs that are used, which indicates that the observed anomalies in the degrees of risk aversion exhibited by the subjects are quite systematic and, as such, can reasonably and convincingly be attributed to the nature of each approach (i.e., the inherently different risk preferences that subjects exhibit with respect to random payoffs and random payoffs).

It is important to note that this result (i.e., PaRA < PrRA) is robust across all the MPL designs in terms of the relative sizes of the elicited degrees of risk aversion (i.e., the average CRRAs computed) across all the three MPL designs. In fact, there is no inconsistency in this finding across the three MPL designs in terms of the average elicited PrRAs being larger than their corresponding PaRAs. However, PrRA is statistically significantly greater than PaRA in two of the designs (H&L and CvU), but PrRA is not statistically significantly greater than PaRA in terms of the Bins. design. Despite this, the average elicited PrRA is still larger than the average elicited PaRA in the Bins. design, but the difference is not big enough to be statistically significant. The degrees of risk aversion elicited under the EUT are somewhat subject to context (the MPL designs in this case), as discussed in the literature of experimental economics (e.g., see Zhou and Hey, 2017 [5]; and Loomes and Pogrebna, 2014 [16]), which could explain why the results of the Bins. design are not statistically significant. As suggested in the literature of experimental economics, the degree of risk aversion should be elicited in the context within which it is supposed to be interpreted. For instance, if probability weighting is a crucial aspect of the real-world setting of interest, researchers should choose the H&L design, while if payoff weighting is a major facet of the real-world phenomenon in the study, then the Bins. design should be followed. More on this matter will be discussed in Section 6.

More formally, the validity and significance of these findings and differences in the degrees of risk aversion can be investigated through statistical hypothesis tests. For the purpose of statistical hypothesis testing, we use a wide range of statistical tests relevant to the research questions of the study, including the Wilcoxon signed-rank test, the Arbuthnott–Snedecor–Cochran sign test, and the two-sample T test for paired data, and the great majority of these statistical tests confirm that **PrRA is statistically significantly greater than PaRA**. Table 2 provides the main results of the above-mentioned statistical tests for each of the three pairs of elicitation procedures. In addition, Appendix E provides all the details of these tests as well as some additional statistical tests.

**Table 2.** Summary of the major results of the three main statistical tests concerning the equivalence of PaRA and PrRA.

| Test | Test Explanation | Test Hypothesis | Task 1 = Task 6 (PaRA = PrRA) | Task 2 = Task 5 (PaRA = PrRA) | Task 3 = Task 4 (PaRA = PrRA) | PaRA = PrRA (Average of Pa and Pr Tasks) | Overall Conclusion |
|---|---|---|---|---|---|---|---|
| **The Wilcoxon matched-pairs signed-rank test** | It tests the equality of matched pairs of observations (non-parametric). | $H_0$: Both distributions are the same. | Reject $H_0$ at 5%  Prob>\|z\|=0.0303  $H_1$: PaRA≠PrRA  Confirm $H_1$ | Reject $H_0$ at 5%  Prob>\|z\|=0.0034  $H_1$: PaRA≠PrRA  Confirm $H_1$ | Fail to reject $H_0$  Prob>\|z\|=0.5619  $H_1$: PaRA≠PrRA  Cannot Confirm $H_1$ | Reject $H_0$ at 5%  Prob>\|z\|=0.0204  $H_1$: PaRA≠PrRA  Confirm $H_1$ | Most of the designs, as well as their average, confirm that **PaRA≠PrRA**. |
| **The Arbuthnott–Snedecor–Cochran sign test** | It tests the equality of matched pairs of observations (non-parametric). | $H_0$: The median of the differences is zero (the true proportion of positive (negative) signs is one-half.) | Reject $H_0$ at 5%  Prob(.) = 0.0407  $H_1$: PaRA < PrRA  Confirm $H_1$ | Reject $H_0$ at 5%  Prob(.) = 0.0178  $H_1$: PaRA < PrRA  Confirm $H_1$ | Fail to Reject $H_0$  Prob(.) = 0.3494  $H_1$: PaRA < PrRA  Cannot Confirm $H_1$ | Reject $H_0$ at 5%  Prob(.) = 0.0110  $H_1$: PaRA < PrRA  Confirm $H_1$ | Most of the designs, as well as their average, confirm that **PaRA < PrRA**. |
| **Two-sample *t* test for paired data** (using midpoint CRRA's) | It tests if two variables have the same mean, assuming paired data (parametric). | $H_0$: The mean of the difference is zero. | Reject $H_0$ at 5%  Prob(T < t) = 0.0147  $H_1$: PaRA < PrRA  Confirm $H_1$ | Reject $H_0$ at 5%  Prob(T < t) = 0.0014  $H_1$: PaRA < PrRA  Confirm $H_1$ | Fail to Reject $H_0$  Prob(T < t) = 0.4306  $H_1$: PaRA < PrRA  Cannot Confirm $H_1$ | Reject $H_0$ at 5%  Prob(T < t) = 0.0092  $H_1$: PaRA < PrRA  Confirm $H_1$ | It shows that most of the designs, as well as their average, confirm that **PaRA < PrRA**. |

Table 2 reports the results of three statistical tests regarding whether the degree of PaRA is equal to the degree of PrRA or whether they are statistically significantly different. The first two tests are non-parametric tests and the last one is a parametric test. The Wilcoxon matched-pairs signed-rank (WMPSR) test is a non-parametric statistical hypothesis test introduced by Wilcoxon (1945) [17] that is most commonly used to test the equality of matched pairs of observations. In fact, it is a paired difference test that compares two related samples, matched samples, or repeated measurements on a single sample to assess whether their population mean ranks differ. The null hypothesis of this test is that both distributions are the same. The p-values associated with this test, which are reported in Table 2, show that most of the MPL designs, as well as their averages, confidently confirm that the degree of PaRA is statistically significantly different from the degree of PrRA. Thus, PaRA and PrRA are not from the same probability distribution, and as such, they should be defined and regarded as two distinct attributes and concepts. As a result, this rejects the hypothesis that the difference between the degrees of PaRA and PrRA is due to chance in this experiment. Hence, on average, the implicit suggestion of the duality theory regarding the equivalence of the degrees of risk aversion elicited from the direct and indirect approach is statistically rejected.

Although the WMPSR test implies that the degrees of PaRA and those of PrRA are statistically significantly different (PaRA ≠ PrRA), it still does not explicitly indicate which one is greater in magnitude. Another similar test which can answer this question is the so-called Arbuthnott–Snedecor–Cochran (ASC) sign test. It is a non-parametric statistical test initially introduced by Arbuthnott (1710) [18], but it was better explained afterwards by Snedecor and Cochran (1989) [19]. This statistical test is typically used to test the equality of matched pairs of observations. The null hypothesis of this test is that the median of the differences is zero, which is, in turn, equivalent to the hypothesis that the true proportion of positive (negative) signs is one-half. One important advantage of the ASC test compared to the WMPSR test is that the ASC test does answer the question of which variable is statistically significantly greater than the other. In other words, not only does the ASC test point out whether or not there is a statistically significant difference, but it also refers to the direction of difference and indicates which variable is greater. As shown in Table 2, as with the results from the WMPSR test, most of the MPL designs, as well as their averages, confidently verify that the degree of PaRA is statistically significantly different from the degree of PrRA, and that PrRA is statistically significantly greater than PaRA. Other results are similar to those from the WMPSR test.

In the two aforementioned statistical tests, no assumptions were made about the underlying distributions. One may want to use a parametric test (e.g., the paired-sample *t* test) as well, and thereby make some assumption about the underlying distributions (e.g., normality). In this case, the relevant test is the so-called two-sample *t* test for paired data (which is the parametric counterpart of the WMPSR non-parametric test). The two-sample *t* test for paired data is a test on the equality of means. In fact, it tests whether or not two variables have the same mean, assuming paired data. In order to perform this parametric statistical test, the mid-point CRRA values exhibited by the subjects are used. (For the first and last implied CRRA intervals, the lower and upper bounds of each interval, respectively, were inevitably used, which fortunately makes the results even more conservative.) The results are fairly commensurate with those of the two previous tests, as reported in Table 2.

A general conclusion that can be drawn from this section is that PrRA is statistically significantly greater than PaRA. This distinction is similar in nature to the well-documented distinction between 'risk aversion' and 'loss aversion', to some extent. Loss aversion, which refers to individuals' general tendency to prefer avoiding losses to attaining equivalent gains, was first identified by Tversky and Kahneman (1979 [20] and 1991 [21], respectively). As Levin et al. (1998) [22, 23] mention, whether a transaction is framed as a loss or as a gain is very important to the elicited degree of aversion. As with the distinction between 'risk aversion' and 'loss aversion', the results of this study show that a transaction being framed with "stochastic payoffs" or "stochastic prices" would be very important to the magnitude of the degree of risk aversion. Thus, the findings of the present paper show that framing a transaction differently (once with payoff odds and another time with price odds) has a significant effect on the degree of risk aversion, consumer behavior, and decision making.

This is an interesting finding that implicitly suggests that individuals, in general, have higher willingness to pay (WTP) for price-guaranteeing insurance premiums than those guaranteeing payoff quantities. It also indicates that risk-preference-related implications of the DT are statistically rejected from a behavioral point of view, since experimental evidence shows that there is a systematic distance from rationality when subjects are exposed to random payoffs versus random prices.

Some scholars including psychologists and neuroscientists (e.g., Tom et al., 2007 [24]; De Martino et al., 2010 [25]; and Canessa et al., 2013 [26]) have studied the structural neural basis of loss aversion in decision-making under risk, based on Kahneman and Tversky's prospect theory. As they report, the reactions in individuals' brains are typically stronger in response to possible losses than to gains. Tom et al. (2007) [24] call this phenomenon neural loss aversion, which provides a neurological explanation for loss aversion. Others have realized that losses may cause greater activity in brain regions that process emotions, including the insula and the amygdala. Within these studies, there are convincing psychological and neurological explanations as to why individuals' degrees of loss aversion are typically greater than that of risk aversion. Likewise, it can be hypothesized that since the stochastic price in this experiment is a buying price at which the subjects buy the widget, this is a price that they have to "pay", which is an amount of money they "lose". In that sense, this price is inherently close to the notion of loss aversion, although not exactly the same; as such, individuals' degrees of aversion with respect to this stochastic price are greater, potentially because of the similar reasons mentioned above for the case of loss aversion, but still smaller than the degree of loss aversion, because it is eventually equivalent to a payoff lottery. After all, developing a clear understanding on how neural processes affect random prices requires further investigation, which are beyond the scope of the present paper.

*5.2. Estimating Payoff Risk Premiums and Price Risk Premiums Using the CvU Design*

The previous sections showed that the degree of PrRA is statistically significantly greater than that of PaRA for the average individual. As discussed earlier, this implies that individuals, in general, have higher willingness to pay (WTP) for price-guaranteeing insurance premiums than those guaranteeing payoff quantities. In turn, this suggests that the risk premium (RP), as a measure of willingness to pay for insuring an uncertain situation, is statistically significantly greater for stochastic prices compared to that for stochastic payoffs. This section aims to investigate the RP differences exhibited by subjects when they are faced with stochastic prices and compare them to when they are faced with stochastic payoffs. For the sake of convenience, in this paper, the former variable is called the price risk premium (PrRP) and the latter variable is called the payoff risk premium (PaRP).

Before defining the RP, it makes sense to first define several other economic concepts that are closely related to the concept of the RP. These concepts include the expected value, the expected utility, and the certainty equivalent, which are usually defined in the context of the EUT and the standard theories of lottery choice. The expected value (EV) of a lottery (in the theory of lottery choice, a lottery in fact represents a random variable) is defined as a measure of the average payoff that the lottery will generate, which, in practice, is the average of the payoffs obtained if one plays the same lottery many times. The expected utility (EU) of a lottery is defined as the EV of the utility levels that a decision maker receives from the potential payoffs of the lottery. In general, the EU of a lottery is lower than the utility of the EV of the lottery for a risk-averse individual, because such an individual has a concave utility function. The certainty equivalent (CE) is defined as the sure (certain) amount of money that leaves an individual indifferent to a lottery. The risk premium (RP) is the difference between the expected payoff of the risky situation (i.e., the EV) and the sure amount for which the risky situation would be traded by an individual (i.e., the CE). Put differently, the RP is the difference between the EV (i.e., the expected payoff) of a lottery and the CE payment (RP = EV − CE). Thus, it is the minimum willingness-to-accept compensation for the risk involved. An RP is positive if the individual is risk-averse.

Among the three MPL designs that are used in this study, the CvU design can directly elicit the two RP measures introduced above (i.e., the RP when faced with stochastic prices and the RP when faced with stochastic payoffs) and compare them with one another. Using the CvU design tasks (i.e., Task 2 and Task 5), one can compute the PaRP and the PrRP exhibited by subjects. As briefly explained earlier, this capability is attained due to the confrontation of certain payoffs and uncertain payoffs that exists within the construction of the CvU design; as a result, the switching point can help us infer the CEs, from which RPs can be readily calculated. Considering the fact that, in the other two MPL designs, there is no such confrontation of certain situations and uncertain situations, the other two designs cannot help us directly elicit RPs. Figures 7 and 8 exhibit histograms and kernel densities of the PaRP (Task 2) and the PrRP (Task 5), respectively, exhibited by the subjects.

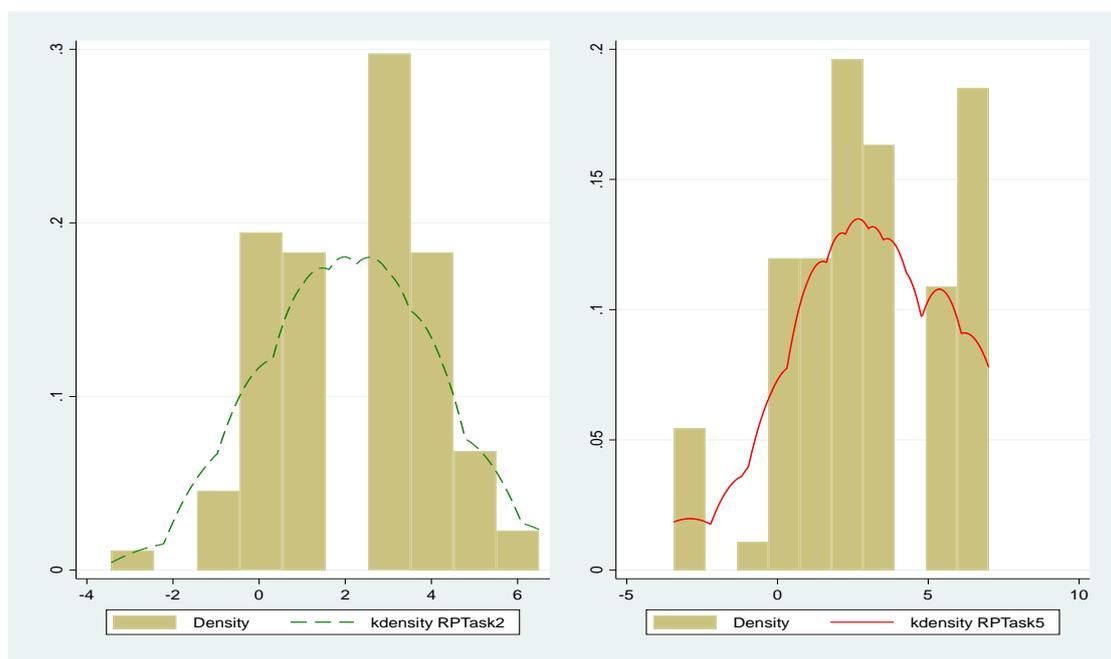

**Figure 7.** Histograms and kernel densities of the PaRP and the PrRP exhibited by the subjects.

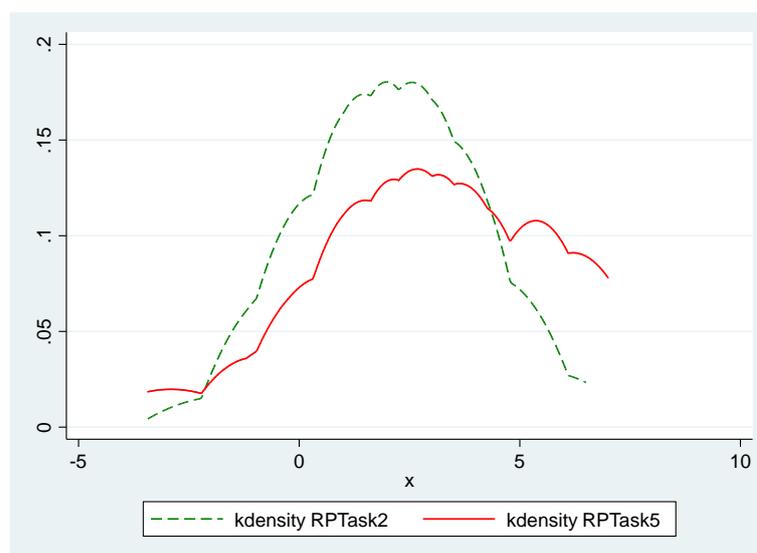

**Figure 8.** Kernel densities of the PaRP and the PrRP exhibited by the subjects.

The results from this section indicate that the RPs inferred from a task with uncertain payoffs are greater than those inferred from a task with uncertain payoffs. More specifically, the average of the PaRPs for the sample under study is USD 2.10 (i.e., **19.1%** of the EV), while the average of the PrRPs is USD 3.05 (i.e., **27.7%** of the EV). This is another reflection of the fact that PaRA < PrRA. Equivalently, it can be said that the CE for lotteries with payoff odds (PaCE) is greater than that for lotteries with price odds (PrCE). In specific, the average of the PaCEs for the sample under study is **USD 8.90**, while the average of the PrCEs is **USD 7.95**.

More formally, we can investigate the validity and significance of these findings and differences in the magnitudes of RPs using formal statistical hypothesis tests, such as the Wilcoxon signed-rank test, the Arbuthnott–Snedecor–Cochran sign test, and the two-sample T test for paired data. As shown in Table 3, all of these statistical tests strongly confirm that the **PrRP is statistically significantly greater than the PaRP**. Table 3 provides the main results of the above-mentioned statistical tests, and Appendix E provides more information on these tests.

Table 3. Summary of the results of three statistical tests on risk premiums.

| Test | Test Explanation | Test Hypothesis | RPTask 2 = RPTask 5 (PaRP = PrRP) | Overall Conclusion |
|---|---|---|---|---|
| **The Wilcoxon matched-pairs signed-rank test** | It tests the equality of matched pairs of observations (non-parametric). | $H_0$: Both distributions are the same. $H_1$: PaRP $\neq$ PrRP | Reject $H_0$ Prob > \|z\| = 0.0014 Confirm $H_1$ | It shows that **PaRP $\neq$ PrRP**. |
| **The Arbuthnott–Snedecor–Cochran sign test** | It tests the equality of matched pairs of observations (non-parametric). | $H_0$: The median of the differences is zero (the true proportion of positive (negative) signs is one-half). $H_1$: PaRP < PrRP | Reject $H_0$ Prob(.) = 0.0178 Confirm $H_1$ | It shows that **PaRP < PrRP**. |
| **The two-sample *t* test for paired data** | It tests if two variables have the same mean, assuming paired data (parametric). | $H_0$: The mean of the difference is zero. $H_1$: PaRP < PrRP | Reject $H_0$ Prob(T < t) = 0.0004 Confirm $H_1$ | It shows that **PaRP < PrRP**. |

Table 3 reports the results of three statistical tests as to whether the magnitude of the PaRP is equal to the magnitude of the PrRP, or whether they are statistically significantly different. The first two tests are non-parametric tests in nature and the last one is a parametric test. The results from all the three tests show that the PaRP is statistically significantly different from the PrRP. Thus, it can be inferred that the PaRP and the PrRP are not from the same probability distribution. More specifically, the results from the last two tests indicate that the PrRP is statistically significantly greater than PaRA, strongly suggesting that individuals, in general, demand a larger premium in order to accept to play a lottery with price odds compared to one with payoff odds. Since the RP for lotteries with payoff odds is smaller than that with price odds, it can be inferred that WTPs used for insurance plans covering uncertain payoffs are smaller than WTPs used for insurance plans covering uncertain prices. This strong experimental evidence base indicates that there is a systematic distance from rationality when subjects are exposed to random payoffs versus random prices. Therefore, behaviorally, individuals are generally more comfortable with accepting uncertain payoffs compared to accepting uncertain prices.

## 6. Summary and Discussion

The elicitation of risk attitudes is an essential element in modern economics. However, much of the attention to risk attitudes has been paid to payoff risk aversion (PaRA) thus far, and the notion of price risk aversion (PrRA) has not received much attention yet.

The duality theory implicitly suggests that the degree of payoff risk aversion (PaRA) must be equivalent to the degree of the price risk aversion (PrRA). This paper tests the accuracy of this theoretical prediction through a lab experiment and uses elicitations that employ payoff-based lottery choices and their equivalent price-based lottery choices. This study uses a laboratory experiment to perform a within-subjects comparison of PaRA and PrRA. Among the most well-known MPL designs, Holt and Laury (2002) [3], Binswanger (1980) [4], and the certainty-versus-uncertainty design (henceforth, H&L, Bins., and CvU designs, respectively) are applied. Six equivalent risk elicitation designs using the above-mentioned MPL designs are adopted and deliberately calibrated in such a way that, given the expected utility theory (EUT) and the duality theory (DT), each should elicit the same degree of risk aversion exhibited by a given rational individual, although the designs differ in form, i.e., in terms of their approaches (i.e., the direct utility function (DUF) vs. the indirect utility function (IUF)) and their MPL designs (i.e., H&L, Bins., and CvU). However, the results show that individuals typically exhibit greater sensitivity to price uncertainty than to equivalent payoff uncertainty when making their decisions.

In the experimental design of this study, attempts were made to address different potential effects to the extent possible, including the "incentive effect", the "income effect", the "wealth effect", the "scale effect", the "endowment effect", the "learning effect", the "order effect", and the "fixed effects", so that any potential differences observed in the results of the six elicitation designs can precisely and accurately be attributed to the phenomena in the study and can answer the research questions of interest in this study. In short, a 3 × 2 design was used, with six elicitation procedures, resulting from three MPL designs (H&L, Bins., and CvU) and two approaches (DUF vs. IUF). Subjects were presented with a menu of choices that permits the degree of risk aversion to be measured, and enabled the experimenter to compare behaviors under uncertainty regarding payoffs and uncertainty about prices. For each of the six treatments, four independent sessions were carried out. Numerous demographic variables were also controlled for, such as age, gender, class status, college, major, financial independency, and family income.

The subjects were students studying at the North Carolina State University. Altogether, 88 students from a range of disciplines participated in the experiments, and the average payoff was USD 16.76 (including a USD 5 participation payment). Each session lasted approximately 75 min, with the first 15–20 min being used for instructions. All the subjects participating in the experiment conducted the tasks using the computers in the experimental economics laboratory of the Department of Economics at North Carolina State University. The popular experimental-economics software zTree was employed for in this lab experiment. In the end, the subjects were paid according to their randomly selected payoffs as they left the experimental laboratory.

The key findings of the study are as follows:

- The **vast majority of subjects are risk-averse**, irrespective of whether the elicitation approach is direct (through the DUF) or indirect (through the IUF). In fact, only a few (less than 5%) of them exhibit risk-loving attitudes, and the rest are either risk-neutral (about 12%) or risk-averse (about 83%), averaged across the six tasks.
- One clear conclusion that emerges from the results of this research study is that individuals exhibit statistically significantly greater degrees of risk aversion when faced with **random prices** (PrRA) compared to when they are faced with **random payoffs** (PaRA). In fact, in all the three pairs of designs, the kernel density representing the degree of PrRA lies to the right of its counterpart that represents the degree of PaRA. This is a remarkable result and a thought-provoking observation. More specifically, the findings indicate that the average of the estimated midpoint CRRAs is equal to 0.597 for PaRA (which implies a '**risk-averse**' attitude), while it is equal to 0.708 for PrRA (which implies a '**very risk-averse**' attitude).
- More interestingly, this result (i.e., **PaRA < PrRA**) is robust across all the three MPL designs that are used, which indicates the fact that the observed anomalies in the degrees of risk aversion exhibited by the subjects are quite systematic; as such, they can sensibly and persuasively be attributed to the nature of each approach (i.e., the inherently different risk preferences that subjects exhibit with respect to random payoffs and random prices). Some scholars have argued that the results of EUT-based elicitations are subject to change with respect to different contextual frameworks. For example, Zhou and Hey (2017) [5] argue that the risk elicitation procedure (i.e., the context) employed in an experiment influences the estimated degrees of risk aversion resulting from the experiment. To account for this possibility and to address this concern, this study considered using three different MPL designs in order to be able to check the sensitivity and robustness of our results to contextual differences. The findings show that the observed anomalies between PaRA and PrRA are systematic; as such, regardless of the MPL designs used, the results of all the designs indicate that PrRA is greater than PaRA. Given the robustness of the results across different contextual designs as well as the existing strong evidence base from different statistical tests, the difference between PaRA and PrRA cannot be attributed to noisy decision making by any means. For the purpose of statistical hypothesis testing, a wide range of relevant statistical tests were used, including the Wilcoxon signed-rank test, the Arbuthnott–Snedecor–Cochran sign test, and the two-sample T test for paired data. The great majority of the above-mentioned statistical tests confirm that **PrRA is statistically significantly greater than PaRA**.

- This implicitly suggests that individuals, in general, have higher willingness to pay (WTP) for price-guaranteeing insurance premiums than those guaranteeing payoff quantities. It also indicates that **risk-preference-related implications of the duality theory (DT) are statistically rejected** from a behavioral point of view, since experimental evidence shows that there is a systematic distance from rationality when subjects are exposed to random payoffs versus random prices.
- In addition, our results suggest that although **"the degrees of risk aversion" elicited under the EUT are somewhat subject to context** (here, the MPL designs) (consistent with the mainstream experimental literature which revealed that context matters when results are produced under the EUT, as discussed by Zhou and Hey (2017) [5]), **the broadly defined "risk attitudes"** (i.e., risk-loving, risk-neutral, and risk-averse) **elicited under the EUT are not subject to as much context**. In fact, 41% of the subjects have exhibited different broadly categorized "risk attitudes" across the six designs (i.e., switching their positions back and forth at least once from risk-loving, to risk-neutral, and/or to risk-averse attitudes), among which 23 percentage points have switched back and forth solely around the narrow border of risk neutrality and risk aversion, meaning that only 18% of the subjects have exhibited all the three different broadly defined "risk attitudes" in their six tasks (even in this case, the results from the vast majority of designs are mostly consistent with, and are close to, each other, and only two of the responses are far from the others).
- Additionally, the results of the study show that the extent of being subject to context is greater for PrRA with a standard deviation of 0.451 for its elicited midpoint CRRAs, compared to PaRA with a standard deviation of 0.381 for its elicited midpoint CRRAs.
- These results imply that the MPL elicitation method, referred to as the context of elicitation in this study, does matter to the estimated "degree of risk aversion", but not much so to the broadly categorized "risk attitudes". Thus, if one's concern is only "the general risk attitude" (that is, the broadly defined risk attitude), there is no need to worry much about the MPL design and context of elicitation. However, if "the degree of risk aversion" is of great concern in their studies, then they need to choose the design that greatly resembles the real-world context of their research study. For example, if probability weighting is a crucial aspect of the real-world setting of interest, researchers should choose the H&L design; if payoff weighting is a major facet of the real-world phenomenon under study, then they should choose to work with the Bins. design; and if the confrontation of certain situations and uncertain situations

- best describes the problem they are studying, then they should select the CvU design. For other general purposes, an average of the three designs may be a good choice that potentially better represents a combination of all the important aspects existing in a more complex real-world situation. Furthermore, it is recommended that researchers and practitioners use multiple designs in their experiments, so that they can gain a better understanding and a more holistic picture of risk aversion characteristics, and also check the robustness of their results with respect to context.
- This finding appears to signal that the degree of risk aversion needs to be elicited in the context within which it is supposed to be interpreted. As Zhou and Hey (2017) [5] state, one should estimate the risk aversion coefficients in its related context, because eliciting these coefficients in an unrelated context could lead to misinterpretations of the data. Additionally, as Loomes and Pogrebna (2014) [16] mention, researchers intending to elicit the degree of risk aversion in their studies should select an elicitation procedure similar to the sort of decisions they are investigating.
- It was also shown that the risk premium (RP), as a measure of willingness to pay for insuring an uncertain situation, is statistically significantly greater for stochastic prices compared to that for stochastic payoffs. The results from all the three statistical tests imply that the PaRP is statistically significantly different from the PrRP. The tests suggest that individuals, in general, demand a larger premium in order to agree playing a lottery with price odds compared to that with payoff odds. This strong experimental evidence base indicates that there is a systematic distance from rationality when subjects are exposed to random payoffs versus random prices. Therefore, behaviorally, we see that individuals are typically more comfortable with uncertain payoffs than uncertain prices.
- Since RPs for lotteries with payoff odds are smaller than those with price odds, it can be inferred that WTPs for price-guaranteeing insurance premiums are greater than those guaranteeing payoff quantities. An implication of this finding is that an insurance company can charge higher premium rates if they frame their insurance coverage around uncertain prices instead of uncertain payoffs, in case there is such a possibility at all in the insurance setting. A fine example of such a situation could be insuring the prices at which farmers buy their inputs or sell their crops (as an indicator of uncertain prices) rather than farmers' yields (as an indicator of their uncertain payoffs). The results of the present study show that the average subject is willing to pay a RP, even as large as 27.7% of the expected value of payoffs when faced with uncertain prices, and 19.1% when faced with uncertain payoffs. This suggests that if an insurer intends to guarantee a level of revenue for an insured farmer through guaranteeing either a crop quantity or a crop price, each of which exposes the company to exactly the same degree of risk, the insurer will be able to charge higher premiums for guaranteeing the price, since the findings of this study imply that subjects have higher levels of risk aversion with respect to a price change compared to its equivalent payoff quantity change. Moreover, as a result of this higher level of risk aversion, they have higher WTP for price-guaranteeing insurance premiums.
- This result suggests that if a benevolent social planner (e.g., the US Federal Crop Insurance Program) intends to provide insurance services to convince risk-averse agents to produce risky products (e.g., farmers to plant risky crops), it would be more effective for the insurance plan to focus on and more efficient for the program budget to be spent on insuring prices rather than yield quantity. This is because agents typically exhibit higher degrees of risk aversion to random prices than their equivalent random payoffs and quantities. As an example of a piece of evidence outside the laboratory, in the real world, such a tendency and high sensitivity to random prices can be noticed by comparing the share of revenue programs (constituting 77% of the volume of the crop insurance policies sold in the US in 2013) and AHP programs (constituting 23% of the volume of the crop insurance policies sold in the US), according to Shields (2015) [27], while the only major difference between the two types of programs is the inclusion of the insurance coverage on yield prices. This result becomes even more interesting if one pays attention to the fact that the historical data have shown that price changes are responsible only for 7% of the losses paid by the US Federal Crop Insurance Program during the period of 2001–2015, according to Good (2017) [28]. However, farmers have still shown a high degree of risk aversion with respect to price changes. After all, it is important to note that making a claim that this difference can be totally attributed to the relatively higher degree of price risk aversion is not reasonable and needs more investigation, since it can be attributed to some other factors as well, such as the relative availability of insurance plans in different areas and for different crops, as well as the relative price of the insurance plans for different areas and for different crops. However, this example is brought up here to introduce some potential examples of real-world evidence that are worth more attention, although this goes beyond the scope of the present paper to attend to these different aspects here in greater detail.

## 7. Conclusions and Further Research

The main contributions of the present paper can be further summarized as follows. This paper shows that PrRA is greater than PaRA, implying that people are more risk-averse to random prices than random payoffs. Additionally, it is shown that the implicit suggestions of the duality theory with respect to risk attitudes are rejected from a behavioral point of view. This study also shows that the risk preferences elicited under the EUT are somewhat subject to context, in the sense that the MPL elicitation method does matter to the estimated "degree of risk aversion", but not much so to the broadly categorized "risk attitudes". Moreover, it is shown that risk premium is statistically significantly greater for stochastic prices compared to that for stochastic payoffs, indicating that there is a systematic distance from rationality when subjects are exposed to random payoffs versus random prices. Most of these results are robust across the different designs and statistical tests that were used.

In this paper, experimental evidence is provided, revealing that individuals typically exhibit greater sensitivity to price uncertainty than to equivalent payoff uncertainty when making decisions. Moving forward with this research, more work needs to be carried out to further investigate the different findings of the present paper. Further research should be undertaken to investigate neural correlates of PrRA and PaRA, while individuals decide what choices to make when faced with different types of lotteries. Research studies have documented that a broad set of areas of the brain (e.g., midbrain dopaminergic regions) show varying functioning and changing activities as potential payoffs and losses are involved. For instance, please see the report by Tom et al. (2007) [24], which studies the neural basis of loss aversion in decision making under risk. Neuroimaging studies and measuring brain activity with functional magnetic resonance imaging (fMRI) to investigate individual variability in PrRA and PaRA can help us better understand the nature of the brain in response to PrRA and PaRA, and can identify specific functional regions within the neural network which explain humans' behavior under uncertainty. Such studies can give great insights and helpful explanations as to how and why PrRA is different from the degree of PaRA.

Another interesting area to focus on for further research is to infer and compare selling price risk aversion (SPrRA) and buying price risk aversion (BPrRA). As Kachelmeier and Shehata (1992) [29] report, there is a significant difference in elicited degrees of risk aversion depending on whether the choice task involves buying or selling contexts. As they elaborate, because subjects typically tend to put a higher selling price on things they own and a lower buying price on things they do not, they indeed exhibit risk-seeking behavior in one case and risk-averse behavior in the other. The author's plan for future research is also to compare the relative size of PrRA and PaRA with the degree of loss aversion. The author also intends to further investigate the relationship between the demographic characteristics of subjects with their elicited degrees of PrRA and PaRA.

Additionally, from a methodological point of view, it is more reasonable and advantageous to conduct research and obtain evidence through various research methods. This study used a lab experiment, while others may want to gather evidence from other empirical research methods based on the relative sizes of PrRA and PaRA. Other methods should be used and their results should be compared in order for scholars to be able to develop a better understanding of contextually different results and find the underlying reasons for the existing behavioral differences.

**Funding:** This research received no external funding.

**Institutional Review Board Statement:** Not applicable.

**Informed Consent Statement:** Informed consent was obtained from all subjects involved in the study.

**Data Availability Statement:** Data used in this article is contained within the appendices and is also publicly accessible at https://zeytoonnejad.wordpress.ncsu.edu/my-research-2/ (accessed on 26 May 2022).

**Conflicts of Interest:** The author declares no conflict of interest.

## Appendix A. Wheel of Duality (WOD) in the consumer theory.

**Figure A1.** Wheel of duality in consumer theory – Source: Moosavian (2016)

**Appendix B. Symbols and notations in the WOD.**

max: maximize
min: minimize
s.t.: subject to
q: vector of quantities consumed
P: vector of prices
M: income
p: vector of normalized prices, i.e., P/M
U(q): direct utility function (or utility function)
$M \geq P.q$: budget constraint
V(P,M): indirect utility function
V(p): indirect utility function with normalized prices
E(P,u): "The" expenditure function
E(P,q): The amount of expenditures
D(q,u): The distance function
$x^M$ (P,M): Marshallian (i.e., uncompensated or Walrasian or ordinary) demand function
$x^M$ (p): vector of normalized Marshallian demand function
p=$\phi$(q): vector of Hotelling-style inverse demand function
$x^C$ (P,u): vector of Hicksian-style (i.e., compensated) demand function
p=$\psi$(q,u): vector of Antonelli-style inverse demand function
H-W Id.: Hotelling–Wold identity
Antonelli: Antonelli equation
Slutsky: Slutsky equation
Roy Id.: Roy's identity
Norm'd Roy Id.: normalized version of Roy's identity
Shephard: Shephard's lemma
Norm'd Shephard: Shephard's lemma with normalized prices
DUF: direct utility function
IUF: indirect utility function
EF: expenditure function
DF: distance function
HIDF: Hotelling-style inverse demand function
MDF: Marshallian demand function
HDF: Hicksian demand function
AIDF: Antonelli-style inverse demand function
EAF: expenditure amount function
BC: budget constrain

**Appendix C. Three tables illustrating the equivalence of the DUF and IUF MPL designs for H&L.**

**Table C1.** *Price-based* **lotteries** along with their corresponding final payoff expected values.

| Option A—(Less Risky) | Option B—(More Risky) | Difference in Final Payoff Expected Values |
| --- | --- | --- |
| 1/10 of **(USD 1.25)**, 9/10 of **(USD 1.56)** | 1/10 of **(USD 0.65)**, 9/10 of **(USD 25.00)** | USD 6.99 |
| 2/10 of **(USD 1.25)**, 8/10 of **(USD 1.56)** | 2/10 of **(USD 0.65)**, 8/10 of **(USD 25.00)** | USD 4.98 |
| 3/10 of **(USD 1.25)**, 7/10 of **(USD 1.56)** | 3/10 of **(USD 0.65)**, 7/10 of **(USD 25.00)** | USD 2.97 |
| 4/10 of **(USD 1.25)**, 6/10 of **(USD 1.56)** | 4/10 of **(USD 0.65)**, 6/10 of **(USD 25.00)** | USD 0.96 |
| 5/10 of **(USD 1.25)**, 5/10 of **(USD 1.56)** | 5/10 of **(USD 0.65)**, 5/10 of **(USD 25.00)** | -USD 1.05 |
| 6/10 of **(USD 1.25)**, 4/10 of **(USD 1.56)** | 6/10 of **(USD 0.65)**, 4/10 of **(USD 25.00)** | -USD 3.06 |
| 7/10 of **(USD 1.25)**, 3/10 of **(USD 1.56)** | 7/10 of **(USD 0.65)**, 3/10 of **(USD 25.00)** | -USD 5.07 |
| 8/10 of **(USD 1.25)**, 2/10 of **(USD 1.56)** | 8/10 of **(USD 0.65)**, 2/10 of **(USD 25.00)** | -USD 7.08 |
| 9/10 of **(USD 1.25)**, 1/10 of **(USD 1.56)** | 9/10 of **(USD 0.65)**, 1/10 of **(USD 25.00)** | -USD 9.09 |
| 10/10 of **(USD 1.25)**, 0/10 of **(USD 1.56)** | 10/10 of **(USD 0.65)**, 0/10 of **(USD 25.00)** | -USD 11.10 |

**Table C2.** Rationally equivalent *widget-based* **lotteries** along with their corresponding final payoff expected values.

| Option A—(Less Risky) | Option B—(More Risky) | Difference in Final Payoff Expected Values |
| --- | --- | --- |
| 1/10 of **(12 units)**, 9/10 of **(9.6 units)** | 1/10 of **(23.05 units)**, 9/10 of **(0.6 units)** | USD 6.99 |
| 2/10 of **(12 units)**, 8/10 of **(9.6 units)** | 2/10 of **(23.05 units)**, 8/10 of **(0.6 units)** | USD 4.98 |
| 3/10 of **(12 units)**, 7/10 of **(9.6 units)** | 3/10 of **(23.05 units)**, 7/10 of **(0.6 units)** | USD 2.97 |
| 4/10 of **(12 units)**, 6/10 of **(9.6 units)** | 4/10 of **(23.05 units)**, 6/10 of **(0.6 units)** | USD 0.96 |
| 5/10 of **(12 units)**, 5/10 of **(9.6 units)** | 5/10 of **(23.05 units)**, 5/10 of **(0.6 units)** | -USD 1.05 |
| 6/10 of **(12 units)**, 4/10 of **(9.6 units)** | 6/10 of **(23.05 units)**, 4/10 of **(0.6 units)** | -USD 3.06 |
| 7/10 of **(12 units)**, 3/10 of **(9.6 units)** | 7/10 of **(23.05 units)**, 3/10 of **(0.6 units)** | -USD 5.07 |
| 8/10 of **(12 units)**, 2/10 of **(9.6 units)** | 8/10 of **(23.05 units)**, 2/10 of **(0.6 units)** | -USD 7.08 |
| 9/10 of **(12 units)**, 1/10 of **(9.6 units)** | 9/10 of **(23.05 units)**, 1/10 of **(0.6 units)** | -USD 9.09 |
| 10/10 of **(12 units)**, 0/10 of **(9.6 units)** | 10/10 of **(23.05 units)**, 0/10 of **(0.6 units)** | -USD 11.10 |

**Table C3.** Rationally equivalent *payoff-based* **lotteries** along with their corresponding final payoff expected values.

| Option A—(Less Risky) | Option B—(More Risky) | Difference in Final Payoff Expected Values |
| --- | --- | --- |
| 1/10 of **(USD 12.00)**, 9/10 of **(USD 9.60)** | 1/10 of **(USD 23.05)**, 9/10 of **(USD 0.60)** | USD 6.99 |
| 2/10 of **(USD 12.00)**, 8/10 of **(USD 9.60)** | 2/10 of **(USD 23.05)**, 8/10 of **(USD 0.60)** | USD 4.98 |
| 3/10 of **(USD 12.00)**, 7/10 of **(USD 9.60)** | 3/10 of **(USD 23.05)**, 7/10 of **(USD 0.60)** | USD 2.97 |
| 4/10 of **(USD 12.00)**, 6/10 of **(USD 9.60)** | 4/10 of **(USD 23.05)**, 6/10 of **(USD 0.60)** | USD 0.96 |
| 5/10 of **(USD 12.00)**, 5/10 of **(USD 9.60)** | 5/10 of **(USD 23.05)**, 5/10 of **(USD 0.60)** | -USD 1.05 |
| 6/10 of **(USD 12.00)**, 4/10 of **(USD 9.60)** | 6/10 of **(USD 23.05)**, 4/10 of **(USD 0.60)** | -USD 3.06 |
| 7/10 of **(USD 12.00)**, 3/10 of **(USD 9.60)** | 7/10 of **(USD 23.05)**, 3/10 of **(USD 0.60)** | -USD 5.07 |
| 8/10 of **(USD 12.00)**, 2/10 of **(USD 9.60)** | 8/10 of **(USD 23.05)**, 2/10 of **(USD 0.60)** | -USD 7.08 |
| 9/10 of **(USD 12.00)**, 1/10 of **(USD 9.60)** | 9/10 of **(USD 23.05)**, 1/10 of **(USD 0.60)** | -USD 9.09 |
| 10/10 of **(USD 12.00)**, 0/10 of **(USD 9.60)** | 10/10 of **(USD 23.05)**, 0/10 of **(USD 0.60)** | -USD 11.10 |

## Appendix D. Additional supplementary tables of results.

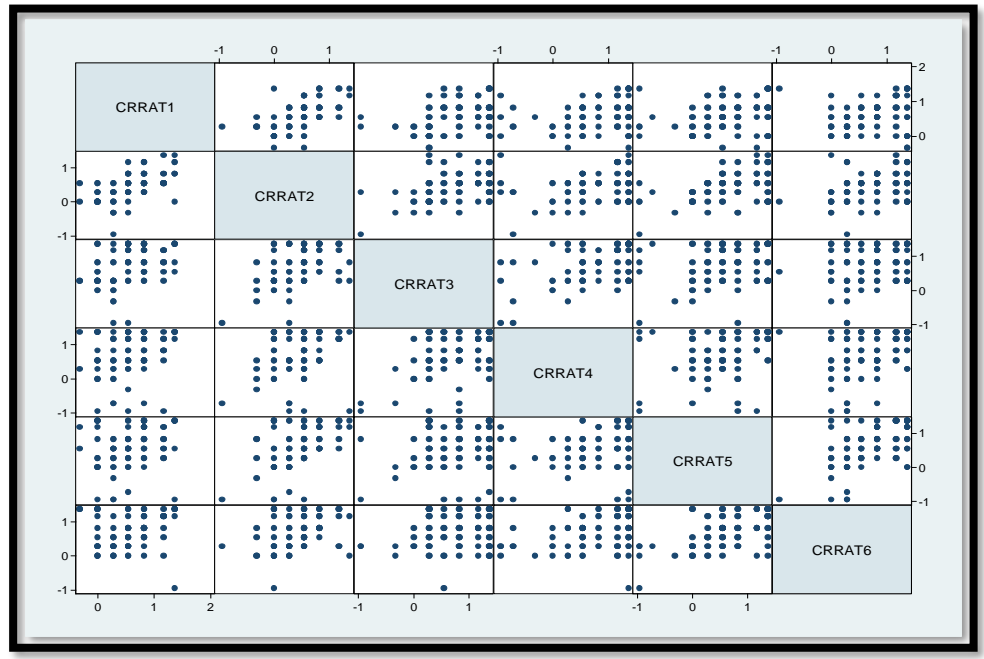

**Figure D1.** The scatter plot matrix of the mid-point CRRA exhibited by the subjects.

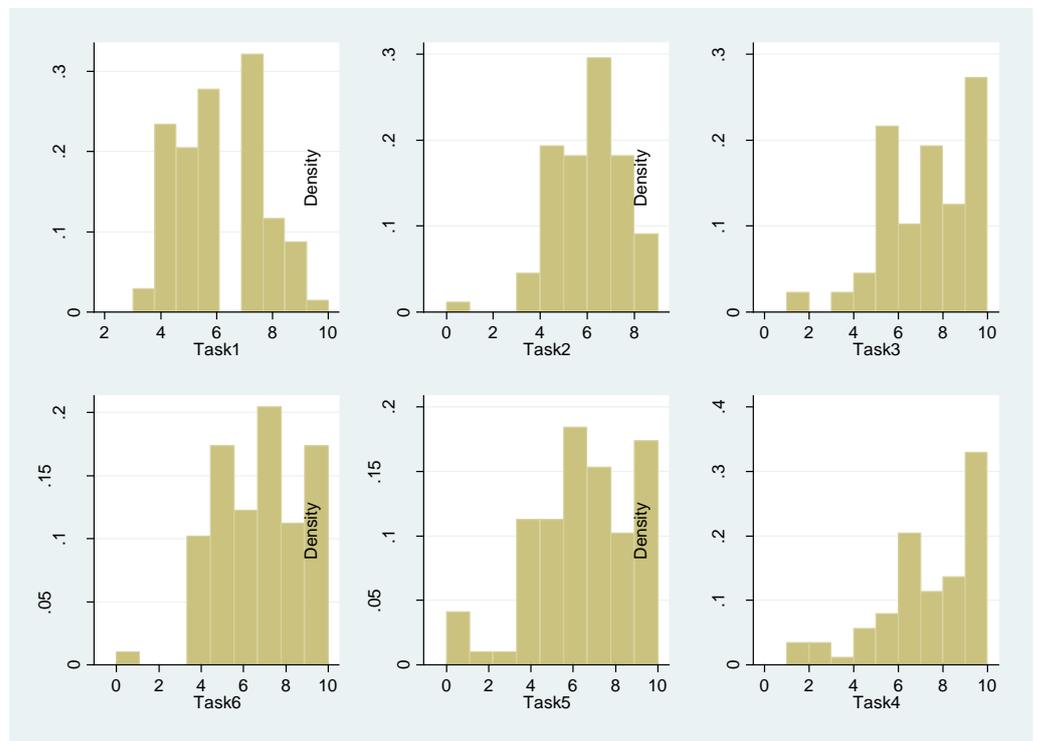

**Figure D2.** Histograms of the switching points and choice numbers. Selected by the subjects, which represent the distributions of the degrees of risk aversion.

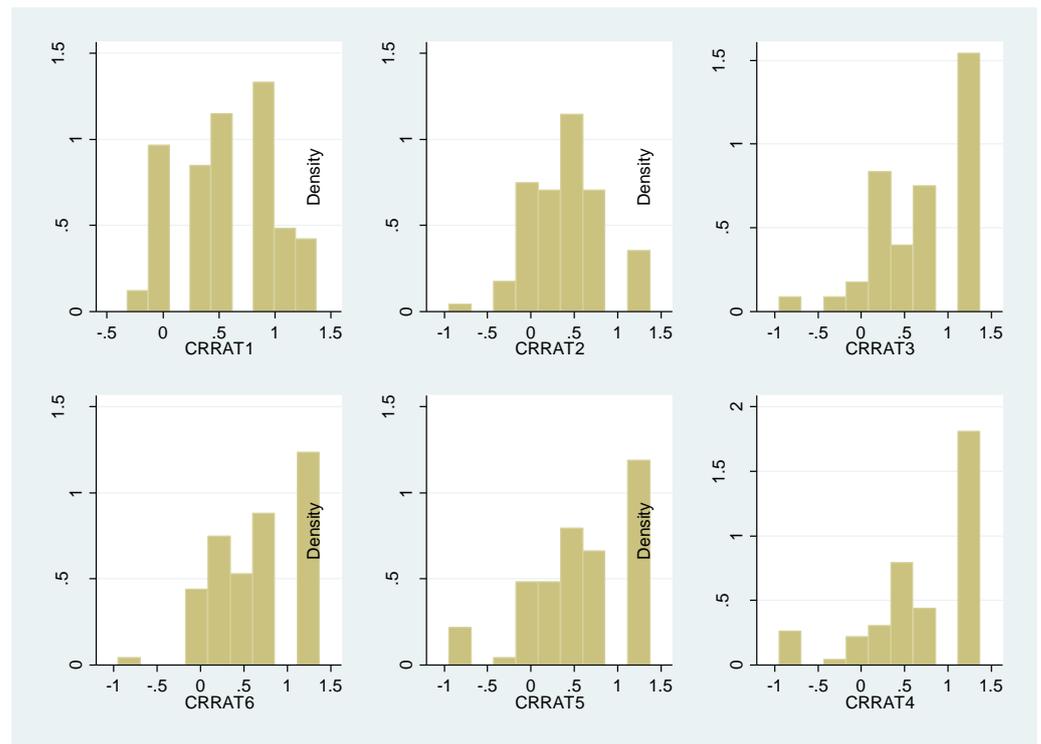

**Figure D3.** Histograms of the CRRA exhibited by the subjects, which represent the distributions of the degrees of risk aversion.

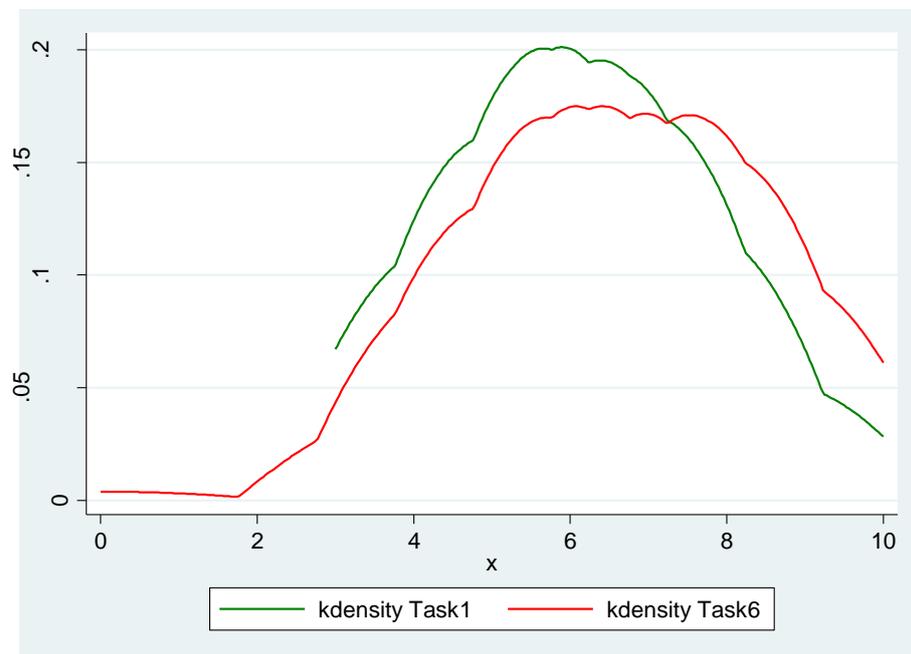

**Figure D4.** Kernel densities of the number of safe choices (i.e., switching points in the H&L tasks). Selected by the subjects, which represent the distributions of the degrees of risk aversion.

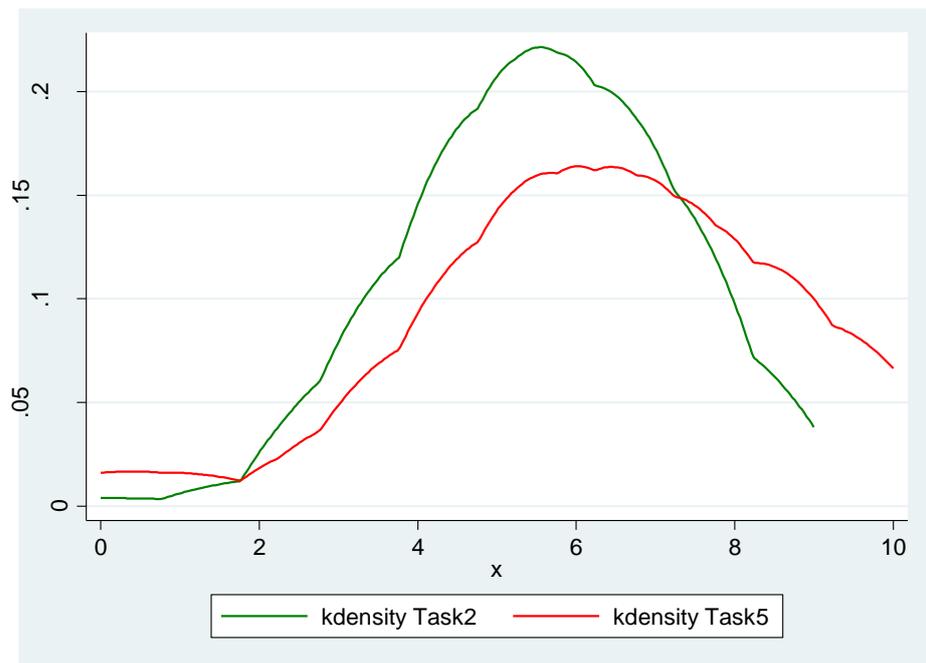

**Figure D5.** Kernel densities of the number of certain choices (i.e., switching points in the CvU tasks). Selected by the subjects, which represent the distributions of the degrees of risk aversion.

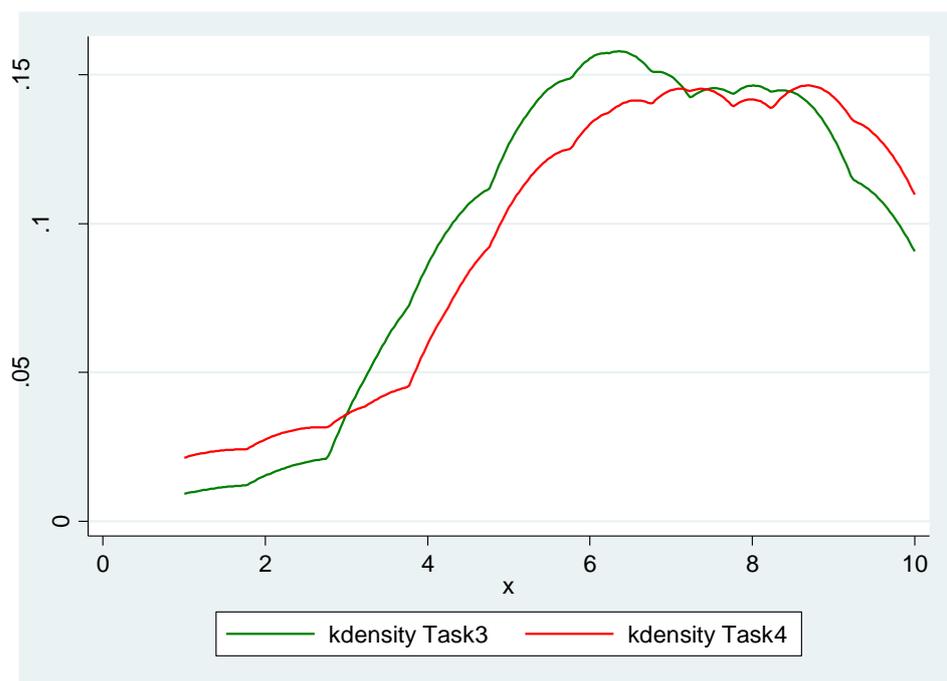

**Figure D6.** Kernel densities of the choice numbers (in Bins. tasks). Selected by the subjects, which represent the distributions of the degrees of risk aversion.

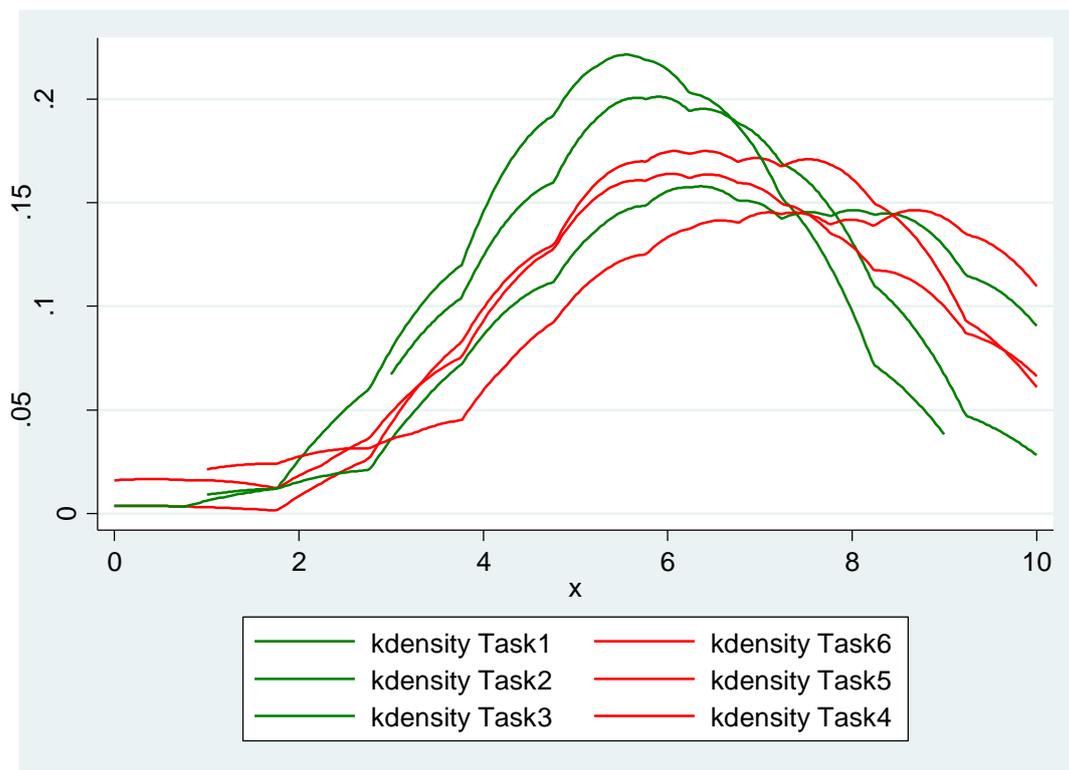

**Figure D7.** Kernel densities of the switching points and choice numbers selected by the subjects, which represent the distributions of the degrees of risk aversion exhibited in direct tasks (Tasks 1-3 shown in green) and those in indirect tasks (Tasks 4-6 shown in red).

**Appendix E. Tables providing the details of statistical hypothesis tests.**

**Table E1.** Results of the Wilcoxon matched-pairs signed-rank test.

| signrank Task1 = Task6 | | | |
|---|---|---|---|
| Wilcoxon signed-rank test | | | |
| **sign** | **obs** | **sum ranks** | **expected** |
| **Positive** | 21 | 1189.5 | 1694 |
| **Negative** | 35 | 2198.5 | 1694 |
| **Zero** | 32 | 528 | 528 |
| **All** | 88 | 3916 | 3916 |
| Unadjusted variance | | 57761 | |
| Adjustment for ties | | −645.88 | |
| Adjustment for zeros | | −2860 | |
| Adjusted variance | | 54,255.13 | |
| Ho: Task1 = Task6 | | | |
| z = −2.166 | | | |
| Prob > z = 0.0303 | | | |

signrank Task 2 = Task 5
Wilcoxon signed-rank test

| sign | obs | sum ranks | expected |
|---|---|---|---|
| **Positive** | 24 | 1138.5 | 1831.5 |
| **Negative** | 42 | 2524.5 | 1831.5 |
| **Zero** | 22 | 253 | 253 |
| **All** | 88 | 3916 | 3916 |

| | |
|---|---|
| Unadjusted variance | 57761 |
| Adjustment for ties | −948.88 |
| Adjustment for zeros | −948.75 |
| Adjusted variance | 55,863.38 |

Ho: Task2 = Task5
$z = -2.932$
$Prob > z = 0.0034$

signrank Task 3 = Task 4
Wilcoxon signed-rank test

| sign | obs | sum ranks | expected |
|---|---|---|---|
| **Positive** | 28 | 1618.5 | 1755 |
| **Negative** | 32 | 1819.5 | 1755 |
| **Zero** | 28 | 406 | 406 |
| **All** | 88 | 3916 | 3916 |

| | |
|---|---|
| Unadjusted variance | 57761 |
| Adjustment for ties | −452.13 |
| Adjustment for zeros | −1928.5 |
| Adjusted variance | 55,380.38 |

Ho: Task 3 = Task 4
$z = -0.580$
$Prob > z = 0.5619$

signrank DUF_Task_Mean = IUF_Task_Mean
Wilcoxon signed-rank test

| sign | obs | sum ranks | expected |
|---|---|---|---|
| **Positive** | 28 | 1368.5 | 1925 |
| **Negative** | 49 | 2481.5 | 1925 |
| **Zero** | 11 | 66 | 66 |
| **All** | 88 | 3916 | 3916 |

| | |
|---|---|
| Unadjusted variance | 57761 |
| Adjustment for ties | 62.25 |
| Adjustment for zeros | −126.5 |
| Adjusted variance | 5752.25 |

Ho: DUF_Task_Mean = IUF_Task_Mean
$z = -2.319$
$Prob > z = 0.0204$

**Table E2.** Results of the Arbuthnott–Snedecor–Cochran sign test.

| signtest Task 1 = Task 6 | | |
|---|---|---|
| Sign test | | |
| sign | observed | expected |
| Positive | 21 | 28 |
| Negative | 35 | 28 |
| Zero | 32 | 32 |
| All | 88 | 88 |

**One-sided tests:**
Ho: median of Task 1 - Task 6 = 0 vs.
Ha: median of Task 1 - Task 6 > 0
Pr(#positive >= 21) =
Binomial(n = 56, x >= 21, $p$ = 0.5) = 0.9780
Ho: median of Task 1 - Task 6 = 0 vs.
Ha: median of Task 1 - Task 6 < 0
Pr(#negative >= 35) =
Binomial(n = 56, x >= 35, $p$ = 0.5) = 0.0407
**Two-sided test:**
Ho: median of Task 1 - Task 6 = 0 vs.
Ha: median of Task 1 - Task 6 = 0
Pr(#positive >= 35 or #negative >= 35) =
min(1, 2*Binomial(n = 56, x >= 35, $p$ = 0.5)) = 0.0814

| signtest Task2 = Task5 | | |
|---|---|---|
| Sign test | | |
| sign | observed | expected |
| Positive | 24 | 33 |
| Negative | 42 | 33 |
| Zero | 22 | 22 |
| All | 88 | 88 |

**One-sided tests:**
Ho: median of Task 2 - Task 5 = 0 vs.
Ha: median of Task 2 - Task 5 > 0
Pr(#positive >= 24) =
Binomial(n = 66, x >= 24, $p$ = 0.5) = 0.9907
Ho: median of Task 2 - Task 5 = 0 vs.
Ha: median of Task 2 - Task 5 < 0
Pr(#negative >= 42) =
Binomial(n = 66, x >= 42, p = 0.5) = 0.0178
**Two-sided test:**
Ho: median of Task 2 - Task 5 = 0 vs.
Ha: median of Task 2 - Task 5 = 0
Pr(#positive >= 42 or #negative >= 42) =
min(1, 2*Binomial(n = 66, x >= 42, $p$ = 0.5)) = 0.0356

| signtest Task3 = Task4 | | |
|---|---|---|
| Sign test | | |
| sign | observed | expected |
| Positive | 28 | 30 |
| Negative | 32 | 30 |
| Zero | 28 | 28 |
| All | 88 | 88 |

**One-sided tests:**
Ho: median of Task 3 - Task 4 = 0 vs.
Ha: median of Task 3 - Task 4 > 0
Pr(#positive >= 28) =
Binomial(n = 60, x >= 28, *p* = 0.5) = 0.7405
Ho: median of Task 3 - Task 4 = 0 vs.
Ha: median of Task 3 - Task 4 < 0
Pr(#negative >= 32) =
Binomial(n = 60, x >= 32, *p* = 0.5) = 0.3494

**Two-sided test:**
Ho: median of Task 3 - Task 4 = 0 vs.
Ha: median of Task 3 - Task 4 = 0
Pr(#positive >= 32 or #negative >= 32) =
min(1, 2*Binomial(n = 60, x >= 32, *p* = 0.5)) = 0.6989

| signtest DUF_Task_Mean = IUF_Task_Mean | | |
|---|---|---|
| Sign test | | |
| sign | observed | Expected |
| Positive | 28 | 38.5 |
| Negative | 49 | 38.5 |
| Zero | 11 | 11 |
| All | 88 | 88 |

**One-sided tests:**
Ho: median of DUF_Task_Mean - IUF_Task_Mean = 0 vs.
Ha: median of DUF_Task_Mean - IUF_Task_Mean > 0
Pr(#positive >= 28) =
Binomial(n = 77, x >= 28, p = 0.5) = 0.9942
Ho: median of DUF_Task_Mean - IUF_Task_Mean = 0 vs.
Ha: median of DUF_Task_Mean - IUF_Task_Mean < 0
Pr(#negative >= 49) =
Binomial(n = 77, x >= 49, *p* = 0.5) = 0.0110

**Two-sided test:**
Ho: median of DUF_Task_Mean - IUF_Task_Mean = 0 vs.
Ha: median of DUF_Task_Mean - IUF_Task_Mean = 0
Pr(#positive >= 49 or #negative >= 49) =
min(1, 2*Binomial(n = 77, x >= 49, p = 0.5)) = 0.022

**Table E3.** Results of two-sample T test for paired data (using mid-point CRRA's).

ttest CRRAT1 = CRRAT6
  Paired *t* test

| Variable | Obs | Mean | Std. Err. | Std. Dec. | [95% Conf. Interval] | |
|---|---|---|---|---|---|---|
| **CRRAT1** | 88 | 0.5765341 | 0.0467863 | 0.4388945 | 0.4835412 | 0.6695269 |
| **CRRAT6** | 88 | 0.7160227 | 0.0526972 | 0.494344 | 0.6112812 | 0.8207642 |
| **diff** | 88 | −0.1394886 | 0.0630229 | 0.5912076 | −0.2647536 | −0.0142237 |

| mean(diff) = mean(CRRAT1−CRRAT6) | | $t = -2.2133$ |
|---|---|---|
| Ho: mean(diff) = 0 | | Degree of freedom = 87 |
| Ha: mean(diff) < 0 | Ha: mean(diff) = 0 | Ha: mean(diff) > 0 |
| Pr(T < t) = 0.0147 | Pr(|T| > |t|) = 0.0295 | Pr(T > t) = 0.9853 |

ttest CRRAT2 = CRRAT5
  Paired *t* test

| Variable | Obs | Mean | Std. Err. | Std. Dec. | [95% Conf. Interval] | |
|---|---|---|---|---|---|---|
| **CRRAT1** | 88 | 0.4475 | 0.0450296 | 0.4224151 | 0.3579988 | 0.5370012 |
| **CRRAT6** | 88 | 0.6297159 | 0.639904 | 0.6002836 | 0.502528 | 0.7569039 |
| **diff** | 88 | −0.1822159 | 0.0591543 | 0.5549168 | −0.2997915 | −0.0646403 |

| mean(diff) = mean(CRRAT2−CRRAT5) | | $t = -3.0803$ |
|---|---|---|
| Ho: mean(diff) = 0 | | Degree of freedom = 87 |
| Ha: mean(diff) < 0 | Ha: mean(diff) = 0 | Ha: mean(diff) > 0 |
| Pr(T < t) = 0.0014 | Pr(|T| > |t|) = 0.0028 | Pr(T > t) = 0.9886 |

ttest CRRAT3 = CRRAT4
  Paired *t* test

| Variable | Obs | Mean | Std. Err. | Std. Dec. | [95% Conf. Interval] | |
|---|---|---|---|---|---|---|
| **CRRAT1** | 88 | 0.7665909 | 0.0584238 | 0.5480637 | 0.6504673 | 0.8827145 |
| **CRRAT6** | 88 | 0.7779545 | 0.0672888 | 0.6312252 | 0.6442107 | 0.9116984 |
| **diff** | 88 | −0.0113636 | 0.0647565 | 0.6074701 | −0.1400743 | 0.117347 |

| mean(diff) = mean(CRRAT3−CRRAT4) | | $t = -0.1755$ |
|---|---|---|
| Ho: mean(diff) = 0 | | Degree of freedom = 87 |
| Ha: mean(diff) < 0 | Ha: mean(diff) = 0 | Ha: mean(diff) > 0 |
| Pr(T < t) = 0.4306 | Pr(|T| > |t|) = 0.8611 | Pr(T > t) = 0.5694 |

ttest DUF_CRRA_Mean = IUF_CRRA_Mean
  Paired *t* test

| Variable | Obs | Mean | Std. Err. | Std. Dec. | [95% Conf. Interval] | |
|---|---|---|---|---|---|---|
| **CRRAT1** | 88 | 0.596875 | 0.0406308 | 0.3811508 | 0.5161169 | 0.6776331 |
| **CRRAT6** | 88 | 0.7078977 | 0.0481272 | 0.4514727 | 0.6122398 | 0.8035557 |
| **diff** | 88 | −0.1110227 | 0.0462402 | 0.4337715 | −0.2997915 | −0.0191153 |

| mean(diff) = mean(DUF_CRRA_Mean−IUF_CRRA_Mean) | | $t = -2.4010$ |
|---|---|---|
| Ho: mean(diff) = 0 | | Degree of freedom = 87 |
| Ha: mean(diff) < 0 | Ha: mean(diff) = 0 | Ha: mean(diff) > 0 |
| Pr(T < t) = 0.0092 | Pr(|T| > |t|) = 0.0185 | Pr(T > t) = 0.9908 |

**Table E4.** Descriptive statistics of CE, RP, and RP/EV.

| sum CETask2 CETask5 | | | | | |
|---|---|---|---|---|---|
| **Variable** | **Obs** | **Mean** | **Std. Dev.** | **Min** | **Max** |
| CETask2 | 88 | 8.899204 | 1.895516 | 4.5 | 14.43 |
| CETask5 | 88 | 7.95375 | 2.735412 | 4 | 14.43 |
| sum RPTask2 RPTask5 | | | | | |
| **Variable** | **Obs** | **Mean** | **Std. Dev.** | **Min** | **Max** |
| RPTask2 | 88 | 2.100796 | 1.895516 | -3.43 | 6.5 |
| RPTask5 | 88 | 3.04625 | 2.735412 | -3.43 | 7 |
| sum RPbyEVTask2 RPbyEVTask5 | | | | | |
| **Variable** | **Obs** | **Mean** | **Std. Dev.** | **Min** | **Max** |
| RPbyEVTask2 | 88 | 19.09814 | 17.23196 | -31.18182 | 59.09091 |
| RPbyEVTask5 | 88 | 27.69318 | 24.86738 | -31.18182 | 63.63636 |

**Table E5.** Results of statistical tests based on the RP (including Wilcoxon matched-pairs signed-rank test, the Arbuthnott–Snedecor–Cochran sign test, and the two-sample T test for paired data).

signrank RPTask2 = RPTask5
Wilcoxon signed-rank test

| **sign** | **obs** | **sum ranks** | **expected** |
|---|---|---|---|
| positive | 24 | 1070.5 | 1831.5 |
| negative | 42 | 2592.5 | 1831.5 |
| zero | 22 | 253 | 253 |
| all | 88 | 3916 | 3916 |

unadjusted variance: 57,761.00
adjustment for ties: −77.00
adjustment for zeros: −948.75
adjusted variance: 56,735.25
Ho: RPTask2 = RPTask5
z = −3.195
Prob > z = 0.0014

signtest RPTask2 = RPTask5
Sign test

| sign | observed | expected |
|---|---|---|
| Positive | 24 | 33 |
| Negative | 42 | 33 |
| Zero | 22 | 22 |
| All | 88 | 88 |

**One-sided tests:**
Ho: median of RPTask 2 - RPTask5 = 0 vs.
Ha: median of RPTask 2 - RPTask5 > 0
Pr(#positive >= 24) =
Binomial(n = 66, x >= 24, *p* = 0.5) = 0.9907
Ho: median of RPTask2 - RPTask5 = 0 vs.
Ha: median of RPTask2 - RPTask5 < 0
Pr(#negative >= 42) =
Binomial(n = 66, x >= 42, *p* = 0.5) = 0.0178
**Two-sided test:**
Ho: median of RPTask2 - RPTask5 = 0 vs.

Ha: median of RPTask2 - RPTask5 = 0
Pr(#positive >= 42 or #negative >= 42) =
min(1, 2*Binomial(n = 66, x >= 42, *p* = 0.5)) = 0.0356

| ttest RPTask2 = RPTask5 | | | | | | |
|---|---|---|---|---|---|---|
| Paired *t* test | | | | | | |
| **Variable** | **Obs** | **Mean** | **Std. Err.** | **Std. Dec.** | **[95% Conf. Interval]** | |
| **CRRAT1** | 88 | 2.100796 | 0.2020626 | 1.895516 | 1.699174 | 2.502417 |
| **CRRAT6** | 88 | 3.04625 | 0.2915958 | 2.735412 | 2.466672 | 3.625828 |
| **diff** | 88 | −0.9454544 | 0.2735494 | 2.56612 | −1.489163 | −0.4017455 |
| mean(diff) = mean(RPTask2−RPTask5) | | | | | t = −3.4562 | |
| Ho: mean(diff) = 0 | | | | | Degree of freedom = 87 | |
| Ha: mean(diff) < 0 | | | Ha: mean(diff) = 0 | | Ha: mean(diff) > 0 | |
| Pr(T < t) = 0.0004 | | | Pr(|T| >|t|) = 0.0008 | | Pr(T > t) = 0.9996 | |

**Appendix F. Data set (only for review purposes). Individual lottery choice decisions—Session A**

**Table F1.** Risk aversion classifications based on chosen options.

| Number of Safe Choices (For HL and CVU Designs) | Selected Decision Number (For Bins. Design) | Range of the Implied Coefficients of RRA for the CRRA Utility Function | Risk Attitude Classifications |
|---|---|---|---|
| 0–1 | 1 | r < –0.95 | Highly risk-loving |
| 2 | 2 | –0.95 < r < –0.49 | Very risk-loving |
| 3 | 3 | –0.49 < r < –0.15 | Risk-loving |
| 4 | 4 | –0.15 < r < 0.15 | Risk-neutral |
| 5 | 5 | 0.15 < r < 0.41 | Slightly risk-averse |
| 6 | 6 | 0.41 < r < 0.68 | Risk-averse |
| 7 | 7 | 0.68 < r < 0.97 | Very risk-averse |
| 8 | 8 | 0.97 < r < 1.37 | Highly risk-averse |
| 9–10 | 9 or 10 | r > 1.37 | Stay in bed (extremely risk-averse) |

| | Tasks with Payoff-Odds | | | Tasks with Price-Odds | | |
|---|---|---|---|---|---|---|
| | Choices Made | | | Choices Made | | |
| Subject | Task1_Choices | Task2_Choices | Task3_Choices | Task4_Choices | Task5_Choices | Task6_Choices |
| A1 | SSSSS / RRRRR | CCCCCC / UUUU | 0000/1/00000 | 000/1/000000 | CCCCC / UUUUU | SSSSS / RRRRR |
| A2 | SSSSSSS / RRR | CCCCCC / UUUU | 000000/1/000 | 00000/1/0000 | CCCCCCC / UUU | SSSSSSS / RRR |
| A3 | SSSSSSSS / RR | CCCCCC / UUUU | 0000000/1/00 | 0000000/1/00 | CCCCCC / UUUU | SSSSSSS / RRR |
| A4 | SSSSS / RRRRR | CCCCC / UUUUU | 000000/1/000 | 0000000/1/00 | CCCCCC / UUUU | SSSSS / RRRRR |
| A5 | SSSSSSS / RRR | CCCCC / UUUUU | 00000000/1/0 | 00000/1/0000 | CCCCC / UUUUU | SSSSSSS / RRR |
| A6 | SSSSSSSS / R | CCCCCCC / UUU | 00000000/1/0 | 000000000/1/ | CCCCCCCC / U | SSSSSSSS / R |
| A7 | SSSSSSS / RRR | CCCCC / UUUUU | 000000000/1/ | 000000000/1/ | CCCCCCCCC / | SSSS / RRRRRR |
| A8 | SSS / RRRRRRR | CCCC / UUUUUU | 0000/1/00000 | 0000/1/00000 | CCCCCC / UUUU | SSSSSSSS / R |
| A9 | SSSS / RRRRRR | CCC / UUUUUUU | 00/1/0000000 | 0000/1/00000 | CCC / UUUUUUU | SSSS / RRRRRR |
| A10 | SSSSSSS / RRR | CCCCCC / UUUU | 000000000/1/ | 0000000/1/00 | CCCCCCCC / U | SSSSSSSS / RR |
| A11 | SSSS / RRRRRR | CCCC / UUUUUU | 0000/1/00000 | 000/1/000000 | CCCC / UUUUUU | SSSS / RRRRRR |
| A12 | SSSS / RRRRRR | CCCCC / UUUUU | 000/1/000000 | 00000/1/0000 | CCCC / UUUUUU | SSSSS / RRRRR |
| A13 | SSSSSSSS / R | CCCCCCC / UUU | 000000000/1/ | 00000000/1/0 | CCCCCCCC / U | SSSSSSSS / R |
| A14 | SSSSSS / RRRR | CCCCCC / UUUU | 000000/1/000 | 000000/1/000 | CCCCCCC / UUU | SSSSSS / RRRR |
| A15 | SSSS / RRRRRR | CCCCC / UUUUU | 000000000/1/ | 000/1/000000 | CCCCCCC / UUU | SSSS / RRRRRR |
| A16 | SSSSSSS / RRR | CCCCCC / UUUU | 0000000/1/00 | 00000/1/0000 | CCCCCC / UUUU | SSSSSS / RRRR |
| A17 | SSSSSSS / RRR | CCCCCCC / UUU | 000000/1/000 | 00000/1/0000 | CCCCCC / UUUU | SSSSSSS / RR |
| A18 | SSSSSS / RRRR | CCC / UUUUUUU | 000000/1/000 | 00/1/0000000 | CCCCC / UUUUU | SSSS / RRRRRR |
| A19 | SSSSSSS / RRR | CCCCCCCC / UU | 00000/1/0000 | 0000000/1/00 | CCCCCCCC / UU | SSSSSSSS / R |
| A20 | SSSS / RRRRRR | CCCC / UUUUUU | 0000/1/00000 | 000000000/1/ | CCCC / UUUUUU | SSSSSSSS / R |
| A21 | SSSSSSSS / RR | CCCCCCCCC / U | 0000/1/00000 | /1/000000000 | CCCCCCCC / UU | SSSS / RRRRRR |
| A22 | SSSSSSSS / R | CCCCCCCCC / U | 000000/1/000 | 00000000/1/0 | CCCCCCCC / U | SSSSSSS / RR |
| A23 | SSSSS / RRRR | CCCCCCC / UUU | 00000000/1/0 | 00000/1/0000 | CCCCCC / UUUU | SSSSS / RRRRR |
| A24 | SSSSS / RRRRR | CCCC / UUUUUU | 0000/1/00000 | 000000/1/000 | CCCCCC / UUUU | SSSSS / RRRRR |
| A25 | SSSSS / RRRRR | CCCC / UUUUUU | 000/1/000000 | 0000000/1/00 | CCCC / UUUUUU | SSSSS / RRRRR |

**Figure F1.** Individual lottery choice decisions—Session A

## F. Individual lottery choice decisions—Session B

**Table F2.** Risk aversion classifications based on chosen options.

| Number of Safe Choices (For HL and CVU Designs) | Selected Decision Number (For Bins. Design) | Range of the Implied Coefficients of RRA for the CRRA Utility Function | Risk Attitude Classifications |
|---|---|---|---|
| 0–1 | 1 | $r < -0.95$ | Highly risk-loving |
| 2 | 2 | $-0.95 < r < -0.49$ | Very risk-loving |
| 3 | 3 | $-0.49 < r < -0.15$ | Risk-loving |
| 4 | 4 | $-0.15 < r < 0.15$ | Risk-neutral |
| 5 | 5 | $0.15 < r < 0.41$ | Slightly risk-averse |
| 6 | 6 | $0.41 < r < 0.68$ | Risk-averse |
| 7 | 7 | $0.68 < r < 0.97$ | Very risk-averse |
| 8 | 8 | $0.97 < r < 1.37$ | Highly risk-averse |
| 9–10 | 9 or 10 | $r > 1.37$ | Stay in bed (extremely risk-averse) |

| | Tasks with Payoff-Odds | | | Tasks with Price-Odds | | |
|---|---|---|---|---|---|---|
| | Choices Made | | | Choices Made | | |
| Subject | Task1_Choices | Task2_Choices | Task3_Choices | Task4_Choices | Task5_Choices | Task6_Choices |
| B1 | SSSSSSSS / RR | CCCCCCC / UUU | 00000/1/0000 | 00000000/1/0 | CCCCCCC / UUU | SSSSSSSS / RR |
| B2 | SSSSSS / RRRR | CCCCCC / UUUU | 00000/1/0000 | 0000/1/00000 | CCCCCC / UUUU | SSSSSSS / RRR |
| B3 | SSSS / RRRRRR | CCCC / UUUUUU | 0000/1/00000 | 000000000/1/ | CCCCCCCCC / | SSSSSSSS / R |
| B4 | SSSSSS / RRRR | CCCC / UUUUUU | 000000000/1/ | 0000000/1/00 | CCCC / UUUUUU | SSSS / RRRRRR |
| B5 | SSSS / RRRRRR | CCCC / UUUUUU | 000000000/1/ | 00000000/1/0 | CCCCCCCCCC / | SSSSSSSS / RR |
| B6 | SSSSSSSS / RR | CCCCCCC / UUU | 0000000/1/00 | 0000000/1/00 | CCCCCC / UUU | SSSSSSSS / RR |
| B7 | SSSSS / RRRRR | CCCCCC / UUUU | 000000/1/000 | 00000000/1/0 | CCCCC / UUUUU | SSSSSS / RRRR |
| B8 | SSSSSSS / RRR | CCCCC / UUUUU | 000000/1/000 | 0/1/00000000 | CCCCCCC / UUU | SSSSSSS / RRR |
| B9 | SSS / RRRRRRR | CCCCCC / UUUU | 0000/1/00000 | 00000000/1/0 | CCCCCCCC / UU | SSSSSSSS / R |
| B10 | SSSSSS / RRRR | CCCCCCC / UUU | 00000000/1/0 | 00000000/1/0 | CCCCCC / UUUU | SSSSSS / RRRR |
| B11 | SSSSSSS / RRR | CCCCCCCC / UU | 0000/1/00000 | 0000000/1/00 | CCCCCCCC / U | SSSSSSSS / RR |
| B12 | SSSSSS / RRRR | CCCCC / UUUUU | 0000/1/00000 | 00000/1/0000 | CCCCCC / UUUU | SSSSSSS / RRR |
| B13 | SSSSSSSS / RR | CCCCCC / UUUU | 000000/1/000 | 000000/1/000 | CCCCCC / UUUU | SSSSSSSS / RR |
| B14 | SSSSSSS / RRR | CCCCCC / UUUU | 000000/1/000 | 000000/1/000 | CCCCCC / UUU | SSSSSSSSS / R |
| B15 | SSSS / RRRRRR | CCCC / UUUUUU | 00000/1/0000 | 00000/1/0000 | CCCCC / UUUUU | SSSS / RRRRRR |
| B16 | SSSSSSS / RRR | CCCCCCC / UUU | 0000000/1/00 | 000000000/1/ | CCCCCCCC / UU | SSSSSSS / RRR |
| B17 | SSSSSSS / RRR | CCCCCC / UUUU | 0000000/1/00 | 0000/1/00000 | CCCCCC / UUU | SSSSSSS / RRR |
| B18 | SSSSSSS / RRR | CCCCCCC / UUU | 00000/1/0000 | 00000/1/0000 | CCCCCC / UUUU | SSSSSSS / RRR |
| B19 | SSSSSS / RRRR | CCCCC / UUUUU | 0000/1/00000 | 00000000/1/0 | CCCCCCCC / U | SSSSSSSS / RR |
| B20 | SSSSS / RRRRR | CCCCC / UUUUU | 000000000/1/ | 0000/1/00000 | CCCCCC / UUUU | SSSS / RRRRRR |
| B21 | SSSSS / RRRRR | CCC / UUUUUUU | 000/1/000000 | 000/1/000000 | CCCCCC / UUU | SSSSSS / RRR |
| B22 | SSSSSSS / RRR | CCCCCCC / UUU | 000000/1/000 | 00000/1/0000 | CCCCCCCCCC / | SSSSSSS / RRR |
| B23 | SSSSSSSSS / R | CCCCCCCC / UU | 00000000/1/0 | 0000000/1/00 | CCCCCCCCCC / | SSSSSSSSS / R |
| B24 | SSSS / RRRRRR | CCCC / UUUUUU | 0000/1/00000 | 00000/1/0000 | CCCCCCCC / U | SSSSS / RRRRR |

**Figure F2.** Individual lottery choice decisions—Session B

## Appendix F. Individual lottery choice decisions—Session C.

**Table F3.** Risk aversion classifications based on chosen options.

| Number of Safe Choices (For HL and CVU Designs) | Selected Decision Number (For Bins. Design) | Range of the Implied Coefficients of RRA for the CRRA Utility Function | Risk Attitude Classifications |
|---|---|---|---|
| 0–1 | 1 | $r < -0.95$ | Highly risk-loving |
| 2 | 2 | $-0.95 < r < -0.49$ | Very risk-loving |
| 3 | 3 | $-0.49 < r < -0.15$ | Risk-loving |
| 4 | 4 | $-0.15 < r < 0.15$ | Risk-neutral |
| 5 | 5 | $0.15 < r < 0.41$ | Slightly risk-averse |
| 6 | 6 | $0.41 < r < 0.68$ | Risk-averse |
| 7 | 7 | $0.68 < r < 0.97$ | Very risk-averse |
| 8 | 8 | $0.97 < r < 1.37$ | Highly risk-averse |
| 9–10 | 9 or 10 | $r > 1.37$ | Stay in bed (extremely risk-averse) |

|  | Tasks with Payoff-Odds ||| Tasks with Price-Odds |||
|---|---|---|---|---|---|---|
|  | Choices Made ||| Choices Made |||
| Subject | Task1_Choices | Task2_Choices | Task3_Choices | Task4_Choices | Task5_Choices | Task6_Choices |
| C1 | SSSSSSSSSS / | CCCC / UUUUUU | 00000/1/0000 | 000000000/1/ | / UUUUUUUUUU | / RRRRRRRRRR |
| C2 | SSSSSSS / RRR | CCCCC / UUUUU | 00000/1/0000 | 000000000/1/ | CCCC / UUUUUU | SSSSSS / RRRR |
| C3 | SSSSSS / RRRR | CCCCC / UUUUU | /1/000000000 | /1/000000000 | CCCCCCC / UUU | SSSSS / RRRRR |
| C4 | SSSSSS / RRRR | CCCCCC / UUUU | 000000000/1/ | 0000/1/00000 | CCCCC / UUUUU | SSSS / RRRRR |
| C5 | SSSSSSS / RRR | CCCCCCCC / UU | 00000000/1/0 | 000000000/1/ | CCCCCC / UUU | SSSSS / RRRRR |
| C6 | SSSSSS / RRRR | CCCCCCC / UUU | 000000000/1/ | 000000000/1/ | CCCCCC / UUUU | SSSSSSS / RRR |
| C7 | SSSSSS / RRRR | CCCCCC / UUUU | 0000000/1/00 | 0000000/1/00 | CCCCCC / UUUU | SSSSSS / RRRR |
| C8 | SSSSSS / RRRR | CCCCCCCC / UU | 0000000/1/00 | 0000000/1/00 | CCCCCC / UUUU | SSSS / RRRRRR |
| C9 | SSSS / RRRRRR | CCCC / UUUUUU | 000/1/000000 | 000/1/000000 | CCCCC / UUUUU | SSSSSS / RRR |
| C10 | SSSSSSS / RRR | CCCCCC / UUUU | 00000/1/0000 | 000000/1/000 | CCCCC / UUUUU | SSSSSS / RRRR |
| C11 | SSSSSS / RRRR | CCCCCCC / UUU | 000000/1/000 | 000000/1/000 | CCCCCCC / UU | SSSSSSS / RRR |
| C12 | SSSSSSSS / RR | CCCCCC / UUUU | 000000/1/000 | 000000/1/000 | CCCCCCCC / UU | SSSSSSSSSS / |
| C13 | SSSSSSSS / RR | CCCCCC / UUUU | 000000/1/000 | 00000/1/0000 | CCCCC / UUUUU | SSSSSS / RRRR |
| C14 | SSSSS / RRRRR | CCC / UUUUUUU | 0000/1/00000 | 0000/1/00000 | CCCCCC / UUU | SSSSSS / RRRR |
| C15 | SSSS / RRRRRR | CCCC / UUUUUU | 0000/1/00000 | 000000000/1/ | CCCCCCCC / UU | SSSSSSSS / R |
| C16 | SSSS / RRRRRR | CCCCC / UUUUU | 000000/1/000 | /1/000000000 | / UUUUUUUUUU | SSSSS / RRRRR |
| C17 | SSSSSSSSS / R | CCCCCCCC / UU | 000000000/1/ | 000000000/1/ | CCCCCCCCCC / | SSSSSSSS / R |
| C18 | SSSSSS / RRRR | CCCCCC / UUUU | 00000000/1/0 | 00000000/1/0 | CCCCCCCCCC / | SSSSSSSS / R |
| C19 | SSSSSS / RRRR | CCCCCC / UUUU | 0000000/1/00 | 0000000/1/00 | CCCCCC / UUU | SSSSSSSS / R |
| C20 | SSSS / RRRRRR | CCCC / UUUUUU | 0000/1/00000 | 000000000/1/ | CCCCCCCCCC / | SSSSSS / RRR |
| C21 | SSSSSSS / RRR | CCCCC / UUUUU | 0000/1/00000 | 00000000/1/0 | CCCCCCCC / U | SSSSSS / RRRR |
| C22 | SSSSSS / RRRR | CCCCC / UUUUU | 00000000/1/0 | 000000000/1/ | CC / UUUUUUUU | SSSSS / RRRRR |

**Figure F3.** Individual lottery choice decisions—Session C

**Appendix F. Individual lottery choice decisions—Session D.**

**Table F4.** Risk aversion classifications based on chosen options.

| Number of Safe Choices (For HL and CVU Designs) | Selected Decision Number (For Bins. Design) | Range of the Implied Coefficients of RRA for the CRRA Utility Function | Risk Attitude Classifications |
|---|---|---|---|
| 0–1 | 1 | $r < -0.95$ | Highly risk-loving |
| 2 | 2 | $-0.95 < r < -0.49$ | Very risk-loving |
| 3 | 3 | $-0.49 < r < -0.15$ | Risk-loving |
| 4 | 4 | $-0.15 < r < 0.15$ | Risk-neutral |
| 5 | 5 | $0.15 < r < 0.41$ | Slightly risk-averse |
| 6 | 6 | $0.41 < r < 0.68$ | Risk-averse |
| 7 | 7 | $0.68 < r < 0.97$ | Very risk-averse |
| 8 | 8 | $0.97 < r < 1.37$ | Highly risk-averse |
| 9–10 | 9 or 10 | $r > 1.37$ | Stay in bed (extremely risk-averse) |

| | Tasks with Payoff-Odds | | | Tasks with Price-Odds | | |
|---|---|---|---|---|---|---|
| | Choices Made | | | Choices Made | | |
| Subject | Task1_Choices | Task2_Choices | Task3_Choices | Task4_Choices | Task5_Choices | Task6_Choices |
| D1 | SSSSS / RRRRR | / UUUUUUUUUU | /1/000000000 | 0/1/00000000 | / UUUUUUUUUU | SSSSS / RRRRR |
| D2 | SSSS / RRRRRR | CCCC / UUUUUU | 00000/1/0000 | 00000/1/0000 | CCCC / UUUUUU | SSSS / RRRRRR |
| D3 | SSSSS / RRRRR | CCCCC / UUUUU | 0000/1/00000 | 00000/1/0000 | CCCCCC / UUUU | SSSSSS / RRRR |
| D4 | SSSSSSSSS / R | CCCCCCC / UUU | 000000000/1/ | 000000000/1/ | CCCCCC / UUUU | SSSSSSSSS / R |
| D5 | SSSSS / RRRRR | CCCCC / UUUUU | 0000/1/00000 | 00000/1/0000 | CCCC / UUUUUU | SSSSSSS / RR |
| D6 | SSSSSS / RRR | CCCCCC / UUUU | 0000000/1/00 | 000000/1/000 | CCCCCCCC / UU | SSSSSS / RRR |
| D7 | SSSS / RRRRRR | CCCC / UUUUUU | 000000000/1/ | 000000000/1/ | CCCCCCCCCC / | SSSSSSSS / R |
| D8 | SSSSSS / RRR | CCCCC / UUUUU | 00000000/1/0 | 00000/1/0000 | CCCC / UUUUUU | SSSSS / RRRRR |
| D9 | SSSS / RRRRRR | CCCCC / UUUUU | 0000/1/00000 | 0000/1/00000 | CCCC / UUUUUU | SSSSSSSSS / R |
| D10 | SSSSSS / RRR | CCCCCCC / UUU | 00000000/1/0 | 000000000/1/ | CCCCCCCC / UU | SSSSSS / RRR |
| D11 | SSSSS / RRRRR | CCCC / UUUUUU | 00/1/0000000 | 0/1/00000000 | CCCC / UUUUUU | SSSSS / RRRRR |
| D12 | SSSSSS / RRR | CCCCCCC / UUU | 000000/1/000 | 00000/1/0000 | CCCCCCC / UUU | SSSSSSS / RRR |
| D13 | SSSSSS / RRR | CCCCCC / UUUU | 00000000/1/0 | 00000000/1/0 | CCCCCCC / UUU | SSSSSS / RRRR |
| D14 | SSSSSSSS / RR | CCCCCC / UUUU | 0000000/1/00 | 000000/1/000 | CCCCC / UUUUU | SSSSSSSS / RR |
| D15 | SSSSSSS / RRR | CCCCCCC / UUU | 00000000/1/0 | 00000000/1/0 | CCCCCCCC / UU | SSSSSSS / RRR |
| D16 | SSSS / RRRRRR | CCCC / UUUUUU | 000000/1/000 | 000000/1/000 | CCCC / UUUUUU | SSSSS / RRRRR |
| D17 | SSSSS / RRRRR | CCCC / UUUUUU | 0000000/1/00 | 0000000/1/00 | / UUUUUUUUUU | SSSS / RRRRRR |

**Figure F4.** Individual lottery choice decisions—Session D

**Appendix F. Data Set (Only for Review Purposes)**

In the tables that follow, the subject number is reported on the left, and demographic data codes are as the following:

A. In what year were you born?
B. What is your gender? **0=Female and 1= Male.**
C. What is your racial or ethnic background? **0=White or Caucasian, 1=Black or African American, 2=Hispanic, 3=Asian, 4=Native American, 5=Multiracial, and 6=Other.**
D. What year in school are you? **0=Freshman, 1=Sophomore, 2=Junior, 3=Senior, 4=Grad Student, and 5=Not Listed.**
E. What is your college? **0=Agriculture and Life Sciences, 1=Design, 2=Education, 3=Poole College of Management, 4=Engineering, 5=Humanities and Social Sciences, 6=Natural Resources, 8=Textiles, 9=College of Veterinary Medicine, 10=Sciences, 11=The Graduate School, and 12=Others.**
F. What is your major? **0=Economics, 1=Business/Management, 2=Accounting, 3=Exploratory Studies, and 4=Others.**
G. How many credit hours are you enrolled in this semester?
H. Not including today, how many previous economics experiments have you participated in? **0, 1, 2, 3, 4, 5, or more(=5).**
I. Not including today, how many previous economics experiments have you participated in where you made repeated choices between lotteries? **0, 1, 2, 3, 4, 5 or more (=5).**
J. What is your marital status? **0=Single and 1=Married.**
K. How many hours do you work in a typical week?
L. How many dollars per hour do you earn in a typical week?
M. Are your financially dependent on your parents? If so, to what extent? **0=Yes, Fully; 1=Yes, Partially; and 2=No, I am independent.**
N. Please indicate the category that best describes your parents' income from all sources before all taxes in 2016. **1=$15,000 and under, 2=$15,001-30,000, 3=$30,001-45,000, 4=$45,001-60,000, 5=$60,001-75,000, 6=$75,001-100,000, and 7=over $100,001.**
O. How many people are in your household? (Yourself and those who live with you and share your income and expenses)
P. What is your weight (in inches)?
Q. What is your height (in pounds)?
R. Please state the country where you were raised.
S. Please state the state where you were raised.
T. **In general**, when it comes to making your economic decisions, which of the following items best describes your risk attitude? **1=Highly risk-loving, 2=Very risk-loving, 3=Risk-loving, 4=Risk neutral, 5=Slightly risk-averse, 6=Risk-averse, 7=Very risk-averse, 8=Highly risk-averse, 9=Stay in bed (Extremely risk-averse).**
U. **In today's experiment**, when making your choices over the lotteries, which of the following items you believe best describes your risk attitude? **1=Highly risk-loving, 2=Very risk-loving, 3=Risk-loving, 4=Risk-neutral, 5=Slightly risk-averse, 6=Risk-averse, 7=Very risk-averse, 8=Highly risk-averse, 9=Stay in bed (Extremely risk-averse).**

**Appendix F. Individual demographic data—Session A.**

**Column Key:** birth year (A), gender (B), race (C), class status (D), college (E), major (F), credit hours (G), # of previous experiments (H), # of previous experiments with risk aversion themes (I), marital status (J), hours worked (K), earnings per hour (L), financial independency (M), family income (N), family member (O), height (P), weight (Q), count®(R), state (S), stated risk attitude toward economic decisions (T), stated risk attitude toward lotteries (U). See data codes on previous pages.

| Subject | A | B | C | D | E | F | G | H | I | J | K | L | M | N | O | P | Q | R | S | T | U |
|---|---|---|---|---|---|---|---|---|---|---|---|---|---|---|---|---|---|---|---|---|---|
| A1 | 1996 | 0 | 0 | 2 | 3 | 1 | 13 | 1 | 0 | 0 | 0 | 0 | 0 | 5 | 1 | 65 | 120 | USA | NC | 5 | 4 |
| A2 | 1998 | 1 | 3 | 0 | 3 | 1 | 16 | 0 | 0 | 0 | 23 | 120 | 1 | 4 | 5 | 67 | 189 | USA | NC | 3 | 3 |
| A3 | 1997 | 1 | 0 | 3 | 4 | 4 | 12 | 1 | 1 | 0 | 0 | 0 | 0 | 6 | 4 | 66 | 150 | USA | NC | 5 | 5 |
| A4 | 1996 | 0 | 0 | 3 | 5 | 4 | 18 | 0 | 0 | 0 | 5 | 15 | 1 | 6 | 1 | 71 | 135 | USA | NC | 3 | 3 |
| A5 | 1998 | 0 | 0 | 1 | 3 | 1 | 13 | 1 | 1 | 0 | 10 | 8 | 1 | 0 | 3 | 68 | 220 | USA | NC | 3 | 3 |
| A6 | 1996 | 0 | 0 | 2 | 3 | 1 | 15 | 1 | 1 | 0 | 0 | 0 | 0 | 4 | 4 | 69 | 125 | USA | NC | 7 | 6 |
| A7 | 1996 | 0 | 0 | 3 | 3 | 1 | 18 | 0 | 0 | 0 | 0 | 0 | 1 | 6 | 4 | 64 | 140 | USA | NC | 6 | 5 |
| A8 | 1999 | 1 | 3 | 0 | 5 | 4 | 15 | 0 | 0 | 0 | 10 | 90 | 0 | 6 | 4 | 67 | 142 | USA | NC | 3 | 3 |
| A9 | 1995 | 0 | 0 | 3 | 0 | 4 | 9 | 2 | 0 | 0 | 8 | 0 | 1 | 3 | 3 | 64 | 122 | USA | NC | 2 | 2 |
| A10 | 1999 | 0 | 1 | 0 | 5 | 4 | 14 | 0 | 0 | 0 | 0 | 0 | 1 | 3 | 3 | 63 | 203 | USA | DE | 4 | 6 |
| A11 | 1998 | 1 | 0 | 1 | 4 | 4 | 17 | 0 | 0 | 0 | 10 | 10 | 1 | 6 | 4 | 71 | 164 | USA | NC | 4 | 4 |
| A12 | 1998 | 1 | 0 | 1 | 5 | 4 | 15 | 0 | 0 | 0 | 0 | 0 | 1 | 5 | 4 | 67 | 132 | USA | NC | 6 | 4 |
| A13 | 1996 | 0 | 0 | 3 | 3 | 1 | 15 | 1 | 1 | 0 | 0 | 0 | 0 | 6 | 4 | 62 | 160 | USA | MA | 4 | 6 |
| A14 | 1995 | 0 | 0 | 3 | 3 | 1 | 17 | 1 | 0 | 0 | 15 | 12 | 1 | 3 | 1 | 70 | 180 | USA | NC | 3 | 3 |
| A15 | 1996 | 0 | 0 | 3 | 8 | 4 | 14 | 0 | 0 | 0 | 5 | 12 | 2 | 6 | 4 | 62 | 140 | USA | IN | 5 | 7 |
| A16 | 1997 | 0 | 1 | 2 | 3 | 1 | 14 | 1 | 0 | 0 | 14 | 9 | 1 | 2 | 3 | 68 | 230 | USA | NC | 5 | 6 |
| A17 | 1999 | 0 | 0 | 0 | 3 | 1 | 17 | 0 | 0 | 0 | 0 | 0 | 0 | 6 | 4 | 54 | 110 | USA | NC | 5 | 4 |
| A18 | 1999 | 1 | 1 | 0 | 3 | 1 | 17 | 0 | 0 | 0 | 0 | 0 | 1 | 5 | 2 | 73 | 206 | USA | NC | 2 | 3 |
| A19 | 1996 | 1 | 5 | 3 | 6 | 4 | 16 | 2 | 1 | 0 | 10 | 80 | 1 | 6 | 5 | 68 | 145 | USA | VA | 3 | 3 |
| A20 | 1997 | 1 | 3 | 1 | 5 | 4 | 12 | 0 | 0 | 0 | 12 | 8 | 1 | 1 | 1 | 67 | 135 | USA | NC | 2 | 3 |
| A21 | 1991 | 1 | 3 | 2 | 5 | 4 | 15 | 0 | 0 | 0 | 20 | 15 | 2 | 1 | 4 | 74 | 235 | USA | CA | 2 | 3 |
| A22 | 1999 | 1 | 2 | 0 | 4 | 4 | 16 | 0 | 0 | 0 | 0 | 0 | 0 | 2 | 4 | 72 | 215 | USA | NC | 3 | 6 |
| A23 | 1995 | 1 | 0 | 3 | 3 | 0 | 14 | 0 | 0 | 0 | 3 | 10 | 0 | 6 | 5 | 70 | 150 | USA | NC | 5 | 5 |
| A24 | 1995 | 1 | 0 | 3 | 5 | 4 | 15 | 1 | 1 | 0 | 25 | 120 | 1 | 6 | 3 | 73 | 155 | USA | NC | 3 | 3 |
| A25 | 1998 | 1 | 0 | 0 | 12 | 3 | 17 | 0 | 0 | 0 | 0 | 0 | 1 | 2 | 3 | 73 | 195 | USA | PA | 3 | 3 |

**Appendix F. Individual demographic data—Session B.**

**Column Key:** birth year (A), gender (B), race (C), class status (D), college (E), major (F), credit hours (G), # of previous experiments (H), # of previous experiments with risk aversion themes (I), marital status (J), hours worked (K), earning per hour (L), financial independency (M), family income (N), family member (O), height (P), weight (Q), country (R), state (S), stated risk attitude toward economic decisions (T), stated risk attitude toward lotteries (U). See data codes on previous pages.

| Subject | A | B | C | D | E | F | G | H | I | J | K | L | M | N | O | P | Q | R | S | T | U |
|---|---|---|---|---|---|---|---|---|---|---|---|---|---|---|---|---|---|---|---|---|---|
| B1 | 1999 | 0 | 0 | 0 | 3 | 2 | 15 | 0 | 0 | 0 | 0 | 0 | 0 | 2 | 3 | 60 | 140 | USA | NC | 6 | 6 |
| B2 | 1998 | 0 | 0 | 1 | 2 | 4 | 16 | 0 | 0 | 0 | 15 | 8 | 1 | 6 | 4 | 66 | 117 | USA | NC | 4 | 3 |
| B3 | 1996 | 1 | 0 | 3 | 3 | 0 | 17 | 0 | 0 | 0 | 8 | 100 | 1 | 6 | 1 | 71 | 145 | USA | NC | 5 | 7 |
| B4 | 1996 | 0 | 0 | 2 | 3 | 1 | 16 | 0 | 0 | 0 | 15 | 9 | 1 | 5 | 4 | 66 | 140 | USA | NC | 6 | 3 |
| B5 | 1997 | 1 | 0 | 2 | 4 | 4 | 16 | 1 | 0 | 0 | 12 | 9 | 0 | 3 | 5 | 75 | 190 | Ireland | County Clare | 5 | 7 |
| B6 | 1995 | 0 | 0 | 3 | 3 | 0 | 3 | 1 | 0 | 0 | 8 | 80 | 0 | 4 | 4 | 66 | 153 | USA | NC | 7 | 7 |
| B7 | 1999 | 1 | 0 | 0 | 4 | 4 | 17 | 0 | 0 | 0 | 0 | 0 | 1 | 6 | 5 | 75 | 185 | USA | NC | 2 | 2 |
| B8 | 1999 | 0 | 2 | 0 | 12 | 3 | 16 | 0 | 0 | 0 | 0 | 0 | 0 | 1 | 6 | 59 | 120 | USA | NC | 5 | 6 |
| B9 | 1996 | 1 | 0 | 3 | 3 | 0 | 18 | 5 | 0 | 0 | 20 | 12 | 0 | 6 | 5 | 72 | 170 | USA | NC | 3 | 3 |
| B10 | 1998 | 1 | 5 | 0 | 3 | 0 | 16 | 0 | 0 | 0 | 0 | 0 | 0 | 5 | 4 | 68 | 144 | USA | NC | 5 | 6 |
| B11 | 1999 | 1 | 2 | 0 | 4 | 4 | 14 | 0 | 0 | 0 | 0 | 0 | 0 | 3 | 3 | 73 | 180 | USA | NC | 4 | 3 |
| B12 | 1996 | 1 | 0 | 2 | 4 | 4 | 12 | 2 | 2 | 0 | 25 | 8 | 1 | 5 | 5 | 71 | 170 | USA | NC | 5 | 5 |
| B13 | 1998 | 0 | 0 | 1 | 0 | 4 | 15 | 0 | 0 | 0 | 6 | 10 | 1 | 6 | 6 | 66 | 125 | USA | MI | 5 | 5 |
| B14 | 1999 | 0 | 0 | 0 | 3 | 1 | 17 | 0 | 0 | 0 | 0 | 0 | 1 | 5 | 2 | 68 | 143 | USA | NJ | 6 | 6 |
| B15 | 1998 | 1 | 0 | 0 | 4 | 4 | 17 | 0 | 0 | 0 | 12 | 9 | 1 | 6 | 4 | 72 | 165 | USA | NC | 4 | 4 |
| B16 | 1998 | 1 | 0 | 1 | 12 | 3 | 14 | 1 | 1 | 0 | 5 | 9 | 0 | 5 | 4 | 69 | 170 | USA | NC | 6 | 7 |
| B17 | 1999 | 1 | 0 | 0 | 5 | 4 | 15 | 0 | 0 | 0 | 0 | 0 | 1 | 5 | 4 | 74 | 152 | USA | NC | 6 | 5 |
| B18 | 1996 | 1 | 2 | 2 | 10 | 4 | 15 | 1 | 1 | 0 | 11 | 10 | 0 | 4 | 3 | 70 | 154 | Bolivia | Santa Cruz de la Sierra | 4 | 4 |
| B19 | 1995 | 1 | 0 | 3 | 3 | 0 | 15 | 2 | 0 | 0 | 20 | 10 | 1 | 6 | 3 | 70 | 200 | USA | NC | 3 | 3 |
| B20 | 1998 | 0 | 3 | 0 | 3 | 1 | 16 | 0 | 0 | 0 | 0 | 0 | 0 | 6 | 4 | 60 | 123 | USA | NC | 3 | 3 |
| B21 | 1996 | 0 | 4 | 3 | 3 | 1 | 14 | 2 | 0 | 0 | 0 | 0 | 1 | 4 | 7 | 66 | 114 | USA | NC | 2 | 3 |
| B22 | 1996 | 1 | 3 | 3 | 4 | 4 | 15 | 1 | 0 | 0 | 4 | 60 | 1 | 6 | 4 | 70 | 145 | USA | NC | 5 | 3 |
| B23 | 1999 | 0 | 0 | 0 | 3 | 0 | 15 | 0 | 0 | 0 | 5 | 10 | 1 | 6 | 3 | 69 | 145 | USA | NC | 7 | 7 |
| B24 | 1996 | 1 | 2 | 3 | 4 | 4 | 15 | 2 | 2 | 0 | 7 | 10 | 0 | 2 | 1 | 70 | 145 | USA | NC | 4 | 5 |

**Appendix F. Individual demographic data—Session C.**

**Column Key:** birth year (A), gender (B), race (C), class status, college (E), major (F), credit hours (G), # of previous experiments (H), # of previous experiments with risk aversion themes (I), marital status (J), hours worked (K), earnings per hour (L), financial independency (M), family income (N), family member (O), height (P), weight (Q), country (R), state (S), stated risk attitude toward economic decisions (T), stated risk attitude toward lotteries (U). See data codes on previous pages.

| Subject | A | B | C | D | E | F | G | H | I | J | K | L | M | N | O | P | Q | R | S | T | U |
|---|---|---|---|---|---|---|---|---|---|---|---|---|---|---|---|---|---|---|---|---|---|
| C1 | 1995 | 0 | 3 | 3 | 4 | 4 | 3 | 1 | 1 | 0 | 0 | 0 | 0 | 3 | 3 | 5 | 120 | China | Jiangxi | 5 | 5 |
| C2 | 1998 | 1 | 0 | 1 | 3 | 2 | 15 | 1 | 0 | 0 | 0 | 0 | 1 | 3 | 4 | 72 | 165 | USA | NC | 6 | 2 |
| C3 | 1998 | 1 | 0 | 0 | 3 | 1 | 14 | 0 | 0 | 0 | 10 | 11 | 1 | 4 | 4 | 70 | 180 | USA | FL | 2 | 2 |
| C4 | 1999 | 1 | 2 | 0 | 4 | 4 | 17 | 0 | 0 | 0 | 30 | 8 | 0 | 2 | 4 | 72 | 170 | USA | NC | 5 | 5 |
| C5 | 1997 | 0 | 3 | 1 | 5 | 4 | 18 | 0 | 0 | 0 | 5 | 8 | 0 | 5 | 2 | 61 | 165 | USA | NC | 7 | 5 |
| C6 | 1990 | 1 | 3 | 4 | 4 | 4 | 3 | 0 | 0 | 0 | 20 | 20 | 1 | 1 | 2 | 70 | 149 | China | Hubei | 3 | 2 |
| C7 | 1998 | 0 | 0 | 0 | 4 | 4 | 18 | 0 | 0 | 0 | 15 | 11 | 1 | 6 | 4 | 65 | 150 | USA | PA | 5 | 5 |
| C8 | 1999 | 0 | 3 | 0 | 3 | 1 | 15 | 0 | 0 | 0 | 0 | 0 | 1 | 4 | 1 | 58 | 114 | USA | NC | 3 | 5 |
| C9 | 1995 | 1 | 0 | 3 | 3 | 2 | 9 | 1 | 0 | 0 | 25 | 10 | 1 | 5 | 8 | 70 | 160 | USA | NC | 4 | 3 |
| C10 | 1999 | 0 | 0 | 0 | 4 | 4 | 16 | 0 | 0 | 0 | 0 | 0 | 0 | 6 | 4 | 64 | 123 | USA | CT | 7 | 6 |
| C11 | 1997 | 0 | 0 | 1 | 3 | 0 | 16 | 0 | 0 | 0 | 0 | 0 | 1 | 6 | 4 | 63 | 120 | USA | NC | 4 | 4 |
| C12 | 1999 | 1 | 0 | 0 | 4 | 4 | 15 | 0 | 0 | 0 | 0 | 0 | 0 | 0 | 3 | 68 | 155 | USA | PA | 6 | 4 |
| C13 | 1997 | 0 | 0 | 2 | 4 | 4 | 16 | 0 | 0 | 0 | 20 | 60 | 1 | 6 | 5 | 65 | 140 | USA | NC | 4 | 5 |
| C14 | 1998 | 1 | 0 | 1 | 4 | 4 | 17 | 1 | 1 | 0 | 0 | 0 | 0 | 6 | 5 | 71 | 150 | USA | MD | 4 | 2 |
| C15 | 1997 | 1 | 0 | 2 | 3 | 1 | 15 | 2 | 0 | 0 | 20 | 40 | 0 | 5 | 6 | 70 | 155 | USA | NC | 5 | 8 |
| C16 | 1999 | 0 | 0 | 0 | 3 | 2 | 16 | 0 | 0 | 0 | 0 | 0 | 2 | 1 | 3 | 70 | 136 | USA | NC | 3 | 2 |
| C17 | 1998 | 1 | 3 | 1 | 3 | 0 | 15 | 0 | 0 | 0 | 10 | 90 | 1 | 0 | 3 | 69 | 150 | USA | NJ | 2 | 7 |
| C18 | 1998 | 0 | 0 | 1 | 3 | 1 | 12 | 0 | 0 | 0 | 16 | 7 | 1 | 4 | 5 | 63 | 125 | USA | NC | 5 | 5 |
| C19 | 1998 | 0 | 2 | 0 | 3 | 1 | 16 | 0 | 0 | 0 | 0 | 0 | 0 | 3 | 5 | 64 | 140 | Venezuela | Anzoategui | 4 | 2 |
| C20 | 1997 | 1 | 0 | 1 | 3 | 0 | 15 | 1 | 0 | 0 | 5 | 5 | 0 | 5 | 4 | 74 | 198 | USA | NC | 3 | 3 |
| C21 | 1999 | 1 | 6 | 0 | 3 | 2 | 17 | 0 | 0 | 0 | 12 | 9 | 1 | 3 | 4 | 72 | 165 | USA | NC | 6 | 6 |
| C22 | 1999 | 1 | 0 | 0 | 3 | 1 | 16 | 1 | 1 | 0 | 0 | 0 | 0 | 5 | 4 | 72 | 135 | USA | NC | 4 | 4 |

**Appendix F. Individual demographic data—Session D.**

**Column Key:** birth year (A), gender (B), race (C), class status (D), college (E), major (F), credit hours (G), # of previous experiments (H), # of previous experiments with risk aversion themes (I), marital status (J), hours worked (K), earning per hour (L), financial independency (M), family income (N), family member (O), height (P), weight (Q), country (R), state (S), stated risk attitude toward economic decisions (T), stated risk attitude toward lotteries (U). See data codes on previous pages.

| Subject | A | B | C | D | E | F | G | H | I | J | K | L | M | N | O | P | Q | R | S | T | U |
|---|---|---|---|---|---|---|---|---|---|---|---|---|---|---|---|---|---|---|---|---|---|
| D1 | 1995 | 1 | 0 | 3 | 3 | 1 | 12 | 3 | 1 | 0 | 16 | 10 | 1 | 1 | 5 | 77 | 225 | USA | NC | 3 | 3 |
| D2 | 1998 | 0 | 0 | 1 | 3 | 0 | 18 | 1 | 0 | 0 | 15 | 10 | 1 | 5 | 4 | 63 | 135 | USA | NC | 5 | 3 |
| D3 | 1996 | 1 | 0 | 3 | 4 | 4 | 12 | 0 | 0 | 0 | 0 | 0 | 1 | 1 | 1 | 72 | 218 | USA | NC | 3 | 3 |
| D4 | 1997 | 1 | 0 | 2 | 3 | 0 | 18 | 1 | 0 | 0 | 20 | 13 | 1 | 5 | 4 | 68 | 150 | USA | DE | 4 | 5 |
| D5 | 1997 | 1 | 0 | 2 | 4 | 4 | 15 | 0 | 0 | 0 | 0 | 10 | 0 | 6 | 6 | 73 | 195 | USA | NC | 5 | 4 |
| D6 | 1998 | 1 | 3 | 1 | 3 | 1 | 16 | 0 | 0 | 0 | 8 | 12 | 0 | 6 | 4 | 75 | 220 | USA | NC | 3 | 5 |
| D7 | 1997 | 1 | 0 | 2 | 3 | 1 | 13 | 1 | 0 | 0 | 5 | 10 | 1 | 6 | 4 | 71 | 185 | USA | NC | 3 | 3 |
| D8 | 1997 | 1 | 3 | 1 | 4 | 4 | 12 | 3 | 0 | 0 | 15 | 8 | 1 | 4 | 3 | 69 | 140 | USA | NC | 4 | 3 |
| D9 | 1996 | 0 | 0 | 3 | 3 | 0 | 11 | 0 | 0 | 0 | 0 | 0 | 1 | 5 | 1 | 64 | 140 | USA | NC | 4 | 3 |
| D10 | 1998 | 1 | 3 | 1 | 4 | 4 | 17 | 0 | 0 | 0 | 7 | 70 | 1 | 6 | 4 | 69 | 120 | USA | NC | 4 | 3 |
| D11 | 1997 | 0 | 0 | 2 | 5 | 4 | 18 | 0 | 0 | 0 | 25 | 12 | 1 | 5 | 6 | 68 | 130 | USA | NC | 2 | 3 |
| D12 | 1999 | 1 | 0 | 0 | 12 | 3 | 16 | 0 | 0 | 0 | 0 | 0 | 0 | 6 | 4 | 74 | 167 | USA | NC | 5 | 4 |
| D13 | 1996 | 0 | 0 | 3 | 0 | 4 | 15 | 1 | 0 | 0 | 12 | 11 | 1 | 6 | 5 | 67 | 175 | USA | NC | 6 | 7 |
| D14 | 1997 | 0 | 1 | 2 | 3 | 1 | 16 | 0 | 0 | 0 | 12 | 50 | 0 | 2 | 4 | 67 | 153 | USA | NC | 5 | 6 |
| D15 | 1997 | 1 | 5 | 2 | 4 | 4 | 14 | 1 | 0 | 0 | 0 | 0 | 1 | 1 | 5 | 66 | 185 | USA | NC | 5 | 6 |
| D16 | 1996 | 1 | 0 | 3 | 4 | 4 | 13 | 0 | 0 | 0 | 4 | 10 | 0 | 6 | 4 | 73 | 185 | USA | NC | 5 | 2 |
| D17 | 1999 | 0 | 0 | 0 | 3 | 1 | 16 | 0 | 0 | 0 | 10 | 10 | 2 | 1 | 6 | 62 | 120 | USA | GA | 2 | 3 |

**Appendix G. Experiment instructions.**

Available at: https://zeytoonnejad.wordpress.ncsu.edu/files/2022/07/Experiment-Instructions.pdf. (Accessed on 26 May 2022)